%% file: Timing.tex
\documentclass[journal,onecolumn,draftcls,12pt]{IEEEtran}
\usepackage[T1]{fontenc}%

\makeatletter
\def\endthebibliography{%
	\def\@noitemerr{\@latex@warning{Empty `thebibliography' environment}}%
	\endlist
}
\makeatother

\IEEEoverridecommandlockouts
\usepackage{cite}
\usepackage{amsmath,amssymb,amsfonts}
\usepackage{algorithm}
\usepackage{algorithmic}
\usepackage{etoolbox}  %
\usepackage{bbm}
\usepackage{amsmath}
\usepackage{tabularx}
\usepackage{amssymb}
\usepackage{threeparttable}
\usepackage{booktabs}
\usepackage{amsthm}
\usepackage{siunitx}

\makeatletter
\let\MYcaption\@makecaption
\makeatother
\usepackage{subcaption}
\makeatletter
\let\@makecaption\MYcaption
\makeatother

\usepackage{bm}
\usepackage{comment}
\usepackage[dvipsnames]{xcolor}
\makeatletter
\patchcmd{\algorithmic}{\addtolength{\ALC@tlm}{\leftmargin} }{\addtolength{\ALC@tlm}{\leftmargin}}{}{}
\makeatother

\makeatletter
\newcommand\fs@betterruled{%
	\def\@fs@cfont{\bfseries}\let\@fs@capt\floatc@ruled
	\def\@fs@pre{\vspace*{5pt}\hrule height.8pt depth0pt \kern2pt}%
	\def\@fs@post{\kern2pt\hrule\relax}%
	\def\@fs@mid{\kern2pt\hrule\kern2pt}%
	\let\@fs@iftopcapt\iftrue}
\floatstyle{betterruled}
\restylefloat{algorithm}
\makeatother

\usepackage{tikz}

\usetikzlibrary{shapes}
\usetikzlibrary{arrows.meta}

\usepackage{pgfplots}
\usepackage{adjustbox}

\usepackage{gincltex}
\usepgfplotslibrary{fillbetween}
\usetikzlibrary{plotmarks}
\usetikzlibrary{patterns}
\usetikzlibrary{calc}

\usepackage{graphicx}
\usepackage{textcomp}
\usepackage{xcolor}
\usepackage{authblk}
\def\BibTeX{{\rm B\kern-.05em{\sc i\kern-.025em b}\kern-.08em
		T\kern-.1667em\lower.7ex\hbox{E}\kern-.125emX}}

\pgfplotsset{compat=1.14}
		
\usepackage{glossaries}

\newacronym{jfi}{JFI}{Jain Fairness Index}
\newacronym{bs}{BS}{Base Station}
\newacronym{ran}{RAN}{Radio Access Network}
\newacronym{mmtc}{mMTC}{massive Machine-Type Communication}
\newacronym{embb}{eMBB}{enhanced Mobile BroadBand}
\newacronym{aoi}{AoI}{Age of Information}
\newacronym{qaoi}{QAoI}{Age of Information at Query}
\newacronym{aoii}{AoII}{Age of Incorrect Information}
\newacronym{voi}{VoI}{Value of Information}
\newacronym{uoi}{UoI}{Urgency of Information}
\newacronym{voip}{VoIP}{Voice over IP}
\newacronym{paoi}{PAoI}{Peak Age of Information}
\newacronym{pdf}{PDF}{Probability Density Function}
\newacronym{mpr}{MPR}{Multi-Packet Reception}
\newacronym{cdf}{CDF}{Cumulative Density Function}
\newacronym{fcfs}{FCFS}{First Come First Serve}
\newacronym{cps}{CPS}{Cyber-Physical System}
\newacronym{fps}{FPS}{frames per second}
\newacronym{qos}{QoS}{Quality of Service}
\newacronym{lcfs}{LCFS}{Last Come First Serve}
\newacronym{haf}{MAF}{Maximum Age First}
\newacronym{aol}{AoL}{Age of Loop}
\newacronym{iot}{IoT}{Internet of Things}
\newacronym{ais}{AIS}{Authentication Identification System}
\newacronym{vdes}{VDES}{VHF Data Exchange System}
\newacronym{leo}{LEO}{Low Earth Orbit}
\newacronym{mec}{MEC}{Mobile Edge Computing}
\newacronym{geo}{GEO}{Geo-synchronous Equatorial Orbit}
\newacronym{meo}{MEO}{Medium Earth Orbit}
\newacronym{e2e}{E2E}{End to End}
\newacronym{m2m}{M2M}{Machine to Machine}
\newacronym{v2v}{V2V}{Vehicle to Vehicle}
\newacronym{v2x}{V2X}{Vehicle to Everything}
\newacronym{ml}{ML}{Machine Learning}
\newacronym{cbr}{CBR}{Constant Bit Rate}
\newacronym{vbr}{VBR}{Variable Bit Rate}
\newacronym{lan}{LAN}{Local Area Network}
\newacronym{vr}{VR}{Virtual Reality}
\newacronym{ar}{AR}{Augmented Reality}
\newacronym{unb}{UNB}{Ultra-Narrowband}
\newacronym{qoe}{QoE}{Quality of Experience}
\newacronym{adsb}{ADS-B}{Automatic Dependent Surveillance - Broadcast}
\newacronym{dl}{DL}{downlink}
\newacronym{ul}{UL}{uplink}
\newacronym{oran}{O-RAN}{Open Radio Access Network Alliance}
\newacronym{csma}{CSMA}{Carrier-Sense Multiple Access}
\newacronym{harq}{HARQ}{hybrid automatic repeat request}
\newacronym{ntp}{NTP}{Network Time Protocol}
\newacronym{tcp}{TCP}{Transmission Control Protocol}
\newacronym{rtp}{RTP}{Real-time Transport Protocol}
\newacronym{tdd}{TDD}{Time-Division Duplex}
\newacronym{rl}{RL}{Reinforcement Learning}
\newacronym{rtt}{RTT}{Round-Trip Time}
\newacronym{ietf}{IETF}{Internet Engineering Task Force}
\newacronym{3gpp}{3GPP}{3rd Generation Partnership Project}
\newacronym{urllc}{URLLC}{Ultra-Reliable Low Latency Communication}
\newacronym{5gacia}{5G-ACIA}{5G Alliance for Connected Industry and Automation}
\newacronym{hl}{HL}{High Layer}
\newacronym{ll}{LL}{Low Layer}
\newacronym{aos}{AoS}{Age of Synchronization}
\newacronym[plural=MDPs,firstplural=Markov Decision Processes (MDPs)]{mdp}{MDP}{Markov Decision Process}
\newacronym{tsn}{TSN}{Time Sensitive Networking}
\newacronym{bsm}{BSM}{Basic Safety Messages}
\newacronym{irsa}{IRSA}{Irregular Repetition Slotted ALOHA}
\newacronym{cam}{CAM}{Cooperative Awareness Messages}
\newacronym{dlt}{DLT}{Distributed Ledger Technology}
\newacronym{sgd}{SGD}{Stochastic Gradient Descent}
\newacronym{fl}{FL}{Federated Learning}
\newacronym{iid}{iid}{independent and identically distributed}
\newacronym{pst}{PST}{Parameter Server Training}
\newacronym{ai}{AI}{Artificial Intelligence}
\newacronym{pca}{PCA}{Principal Component Analysis}
\newacronym{ttff}{TTFF}{Time To First Fix}
\newacronym{sdma}{SDMA}{Space Division Multiple Access}
\newacronym{mse}{MSE}{Mean-Square Error}

\definecolor{violet}{rgb}{0.6,0,0.6}%
\definecolor{orange_D}{rgb}{1,0.3,0}%
\definecolor{cyan}{rgb}{0,0.67,0.64}%
\definecolor{red}{rgb}{0.9,0,0}%
\definecolor{green}{rgb}{0,0.6,0}
\definecolor{yellow}{rgb}{1,0.8,0}

\begin{document}

\title{A Perspective on Time towards Wireless 6G}

\author{Petar Popovski,~\IEEEmembership{Fellow,~IEEE,} 
Federico Chiariotti,~\IEEEmembership{Member,~IEEE,} Kaibin Huang,~\IEEEmembership{Fellow,~IEEE,}
Anders E. Kal\o r,~\IEEEmembership{Student Member,~IEEE,} 
Marios Kountouris,~\IEEEmembership{Senior Member,~IEEE,} 
Nikolaos Pappas,~\IEEEmembership{Member,~IEEE,} and 
Beatriz Soret,~\IEEEmembership{Member,~IEEE} \thanks{P. Popovski (corresponding author, email: petarp@es.aau.dk), F. Chiariotti (fchi@es.aau.dk), A.E. Kal\o r (aek@es.aau.dk) and B. Soret (bsa@es.aau.dk) are with the Department of Electronic Systems, Aalborg University, Denmark. K. Huang (huangkb@eee.hku.hk) is with the Department of Electrical and Electronic Engineering, Honk Kong University. M. Kountouris (marios.kountouris@eurecom.fr) is with the 
Communication Systems Department, EURECOM, France. N. Pappas (nikolaos.pappas@liu.se) is with the Department of Science and Technology, Link\"oping University, Sweden.
}
}
\maketitle

\begin{abstract}
With the advent of 5G technology, the notion of \emph{latency} got a prominent role in wireless connectivity, serving as a proxy term for addressing the requirements for real-time communication. As wireless systems evolve towards 6G, the ambition to immerse the digital into physical reality will increase. Besides making the real-time requirements more stringent, this immersion will bring the notions of time, simultaneity, presence, and causality to a new level of complexity. A growing body of research points out that latency is insufficient to parameterize all real-time requirements. Notably, one such requirement that received a significant attention is information freshness, defined through the \emph{\gls{aoi}} and its derivatives. In general, the metrics derived from a conventional black-box approach to communication network design are not representative for new distributed paradigms such as sensing, learning, or distributed consensus. The objective of this article is to investigate the general notion of timing in wireless communication systems and networks and its relation to effective information generation, processing, transmission, and reconstruction at the senders and receivers. We establish a general statistical framework of \emph{timing} requirements in wireless communication systems, which subsumes both latency and \gls{aoi}. The framework is made by associating a timing component with the two basic statistical operations, decision and estimation. We first use the framework to present a representative sample of the existing works that deal with timing in wireless communication. Next, it is shown how the framework can be used with different communication models of increasing complexity, starting from the basic Shannon one-way communication model and arriving to communication models for consensus, distributed learning, and inference. Overall, this paper fills an important gap in the literature by providing a systematic treatment of various timing measures in wireless communication and sets the basis for design and optimization for the next-generation real-time systems. 
\end{abstract}

\IEEEpeerreviewmaketitle
\glsresetall

\section{Introduction}
How soon is \emph{now}? When two events occur simultaneously? These seemingly na\"{i}ve questions have led to fundamental shifts in physics through the theory of relativity and irrevocably altered our notion of time. Besides the physical time, in a system with various interacting components what matters is the \emph{perception of time}. This is succinctly illustrated by the following excerpt from the novel ``Recursion'' by Blake Crouch~\cite{CrouchBlake2019}: 

\begin{quote}
  \emph{``Just what your brain does to interpret a simple stimulus like that is incredible. The visual and auditory information arrive at your eyes and ears at different speeds, and then are processed by your brain at different speeds. Your brain waits for the slowest bit of stimulus to be processed, then reorders the neural inputs correctly, and lets you experience them together, as a simultaneous event -- about half a second after what actually happened. We think we're perceiving the world directly and immediately, but everything we experience is this carefully edited, tape-delayed reconstruction.''}
\end{quote}

The perception of time, simultaneity, synchronicity, causality -- all these notions get to a new level of complexity as wireless communication offers remote interaction among humans and machines over extended distances. 
Indeed, wireless communication technology is radically transforming the very nature of human interactions, having a profound impact across our society and economy. Various names are used, such as Tactile Internet or Internet of Senses~\cite{simsek20165g}, to denote the trend in which wireless connectivity augments human capabilities beyond their natural domain, enabling operation and interaction with objects and subjects placed within an extended space-time domain. We are at the dawn of the era of connected intelligence and {autonomous} automation, in which a myriad of interconnected sensor-empowered devices with computing, learning, and decision-making capabilities will underpin the global functioning of our societies, enabling formidable progress at industrial, health, transportation, environmental, and educational sectors. 

Naturally, these interactive applications need to perform a series of actions to work, all of which require some time: these include both the actual communication of the necessary data and the computation at both ends, e.g., to compress the raw sensor data into a more compact version on one side, then decode it and present it to the user on the other. These different components make up a latency budget~\cite{alfadhli2019latency,elbamby2018toward}, which must satisfy strict requirements to maintain the real-time illusion. Just as humans collect and process stimuli in the brain, the processor of a device or robot gathers data from its sensors (including communication from other devices) and uses algorithms to create a unified estimate of the environment and translate data from the physical world into the digital domain~\cite{viswanathan2020communications}. In a more general sense, we can talk about interactivity and a ``real-time illusion'' not just for humans, but also for machines: the perception of sensors and the granularity in time of control algorithms and actuators depend on the limitations of the hardware and software, and any timing difference shorter than this granularity will be unnoticeable to them. Naturally, while the latency budget in human communications has a hard floor given by the limits of perception, the latency budget for machines will depend on the specific device and its capabilities, as well as on the application for which it is used.

In order to keep the perception of time close to how ``now'' is commonly defined for all potential real-time applications, latency has been heralded as one of the main features of the widely publicized 5G wireless systems, as well as of the wireless systems beyond 5G. One of the three generic 5G services is \gls{urllc}~\cite{popovski2018URLLC,bennis2018ultrareliable}, where the ambition is to guarantee with very high reliability (e.g., $>99.999$\%) that a given data packet will be delivered within a very short time frame (e.g., on the order of $1$~ms). In a sense, \gls{urllc} aims to satisfy the least common denominator in terms of latency for all potential applications that exist or will emerge in the future. The upshot is that a wireless \gls{urllc} link cuts a small, predictable part of the latency budget in the overall digital service or application. Thus, if the end application has a more relaxed latency budget, then \gls{urllc} creates a higher flexibility in designing the other system modules, such as compression or computation. Most of the research on low latency communications in 6G is following the same pattern, putting even stricter constraints on latency and reliability and foreseeing the use of new technologies, such as terahertz communications and intelligent surfaces, to meet them~\cite{hassan2021key}. While 6G will be the first generation of mobile networks to natively rely on learning optimization at all layers~\cite{she2021tutorial} for resource allocation, parameter optimization, and network orchestration~\cite{wei2020intent}, most proposed solutions take the existing classes of traffic for granted~\cite{salh2021survey}, explicitly using the requirements of \gls{urllc} as their main objectives. The requirements for new use cases and applications, such as \gls{vr} or tactile communications~\cite{giordani2020toward}, are also formulated in terms of the existing classes of traffic, with the same broad objectives as 5G.

\begin{figure}[t]
    \centering
    \input{./figures/basic_scenario.tex}
    \caption{A simple scenario in which two users of different types transmit in the uplink to a common Base Station (BS). One is a high rate user and the other user sends intermittent critical updates. (a) Scheme with low latency reservations for the intermittent user. (b) Scheme with pull-based updates from the intermittent user.}
    \label{fig:HighRatevsLatency}
\end{figure}
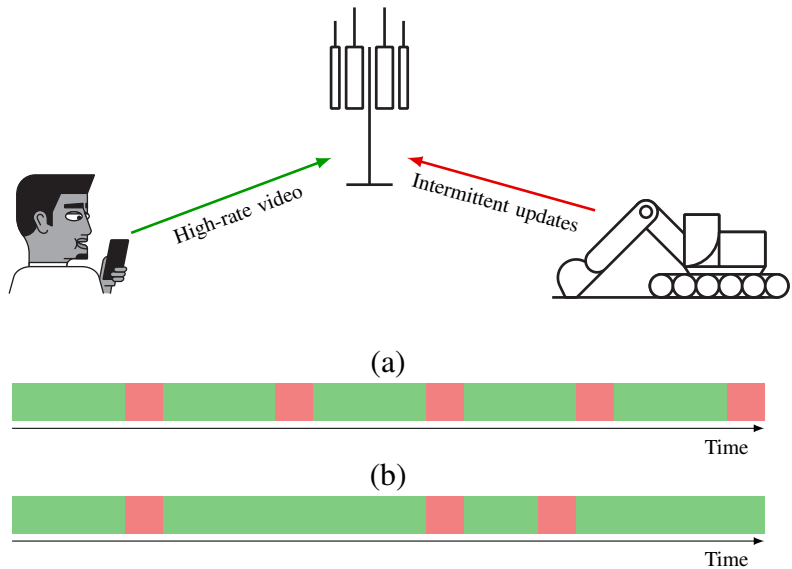

Nevertheless, a growing body of emerging scenarios and research results points out that the maximalistic approach of \gls{urllc} to time-sensitive applications and services is limited. To illustrate this claim, consider the simple scenario on Fig.~\ref{fig:HighRatevsLatency}, where two different users share an uplink channel to a common \gls{bs}. One of the users requires a high rate, while the other sends intermittent critical updates as a part of a networked control system. In a practical setting, the same \gls{bs} may not be the communication end-point for both users, so here we assume that the \gls{bs} has an edge server for the low latency user, while the high-rate user may have a different end-point and the timing aspects for this user are irrelevant in this discussion. Yet, the model is sufficient to show the interactions among the two services over the access resources. Fig.~\ref{fig:HighRatevsLatency}a shows the case that reflects a conservative \gls{urllc} approach, in which low latency transmissions have reserved slots that guarantee maximal latency of at most four time slots. As mentioned above, this can make the system less efficient when the intermittent user has no updates in certain slots, since those slots remain unused by the high rate user. Fig.~\ref{fig:HighRatevsLatency}b shows the same setting, but with the assumption that the transmission of the updates from the intermittent user is based on a \emph{pull-based communication model}~\cite{holm2021freshness}. To elaborate, the \gls{bs} features a predictive controller that can estimate in advance when it will need the next update, based on the freshness of the current update and the state of the system. Naturally, this prediction will also depend on the variability and the temporal evolution of the physical phenomenon observed by the monitoring control device. This simple example illustrates how \emph{predictability} can improve system efficiency, while satisfying the timing objectives without strict reservations defined by the minimal latency.

Based on this example, one can extrapolate more general conclusions about the insufficiency of the latency-only focused design of \gls{urllc}: 
\begin{itemize}
    \item In a \emph{quantitative} sense, aiming always for the least common denominator is inefficient due to the fact that low latency requires resource allocation that may lead to over-provisioning, thereby preventing other services from using the communication resources.
    \item In a \emph{qualitative} sense, there are multiple clear indications that the timing relations in a communication system cannot be condensed solely in the measure of communication latency. This is best illustrated by the emergence of alternative measures of timing, such as \emph{\gls{aoi}}~\cite{kaul2012real, kosta2017age} which aims to quantify the freshness of the data updates coming from, e.g., an \gls{iot} sensor. However, these are only instances of a general measure of timing in a distributed system with communication links. For instance, latency is usually measured with respect to a fixed point in time and space at which a data packet has been created, while \gls{aoi} is measured with respect to the physical state of a certain \gls{cps} or the occurrence of an event. In a distributed system of interconnected nodes, there could be other, more complex notions of timing or latency related to, for example, consensus or a distributed decision process. 
\end{itemize}

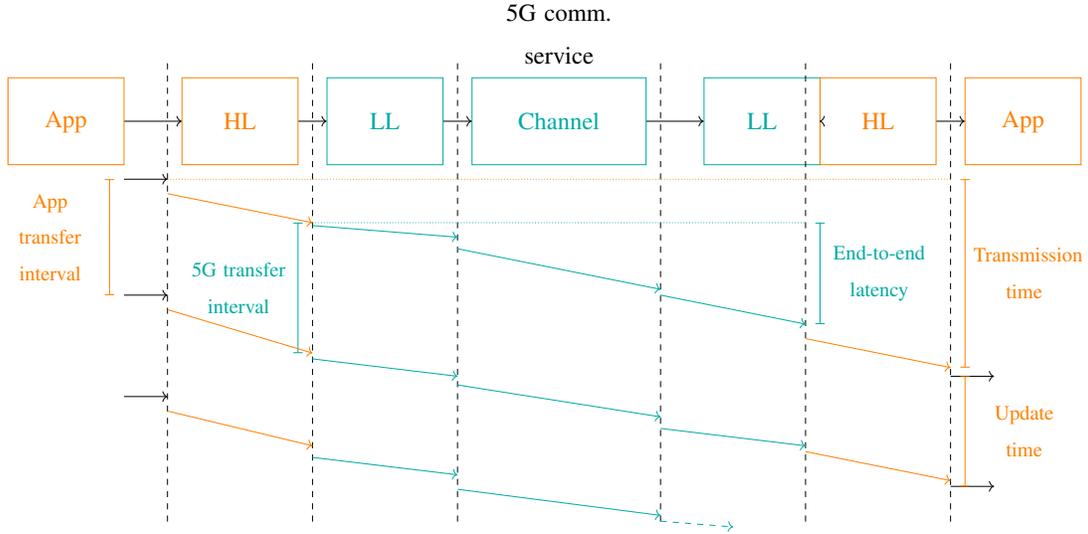
\begin{figure}[t]
    \centering
    \resizebox{0.9\linewidth}{!}{
    \input{figures/timing_model}}\vspace{0.3cm}
    \caption{General timing model from 3GPP~\cite{3GPPTS22.104}.}
    \label{fig:timing_model}
\end{figure}

Fig.~\ref{fig:timing_model} is adapted from a recent \gls{3gpp} technical document~\cite{3GPPTS22.104}. It shows that the standardization of mobile networks is also considering a wider approach to timing issues rather than simply setting extremely strict constraints on the wireless segment. Considering the timing of the application itself, as well as the higher communication layers, can lead to a better resource efficiency, and considering different metrics such as \gls{aoi} can increase the flexibility of the system with respect to the needs of different applications.

The objective of this article is to investigate the general notion of timing in wireless communication systems and networks and its relation to effective information generation, processing, transmission, and reconstruction at the senders and receivers. We provide a systematic introduction to different timing measures, through the way these measures interact with the layers in increasingly complex communication models. In this emerging heterogeneous, and often distributed, networking ecosystem, a general definition of the optimal communication system can be the one that chooses or generates the right piece of information that has to be efficiently transmitted at the right time instant, typically to achieve specific goals. Although the timing measures can be applied to general communication systems and networks, as also illustrated by treating the problems of consensus and distributed learning, the discussion in this article is biased towards the wireless access part. In practical systems, the latency of the core network represents a significant portion of the timing budget. However, the framework introduced in this paper can accommodate the latency of the core network, as exemplified in the model with cascaded modules on Fig.~\ref{fig:timing_model} and Fig.~\ref{fig:CascadedModules}.

We argue that the definition of \emph{the right time instant} is not universal, and that conventional approaches and metrics do not satisfy {the requirements of} many current and future applications and communication networks. Under this perspective, the fundamental problem of communication becomes that of reconstructing the information generated at a source space-time point in a way that is sufficiently accurate for achieving a specific goal in a timely and effective manner at another, target space-time point. Furthermore, in specific scenarios, such as distributed learning, the communication system includes the post-processing of the received information in order to achieve a certain goal.

\subsection{Contributions and Paper Organization}

The main contributions of the paper can be summarized in the following list:
\begin{itemize}
    \item Establish the context for defining timing measures in communication systems by considering the impact of various factors, such as the actors in the given communication scenario or the considered timing scales. 
    \item Provide a comprehensive view and systematization of the timing measures used in the research literature as well as in standardization. 
    \item Define a statistical framework for timing that is sufficiently general to encompass {possibly} all timing measures discussed in the literature. 
    \item Show how the statistical framework for timing can be applied in the context of different communication models and how a given timing measure can be optimized within a given communication model. 
\end{itemize}

The paper is organized as follows. First, we introduce the context for timing measures and describe the main use cases and communication actors in Sec.~\ref{sec:definition}. Our statistical framework, based on the concept of different timing references, is given in Sec.~\ref{sec:framework}. In  Sec.~\ref{sec:sota}, the framework is used to describe the current state of the art on timing, including latency, deadlines, and \gls{aoi}. We then present in Sec.~\ref{sec:p2p} the use of our framework in single-connection communication models, including Shannon's one-way communication model, two-way links, and connection through a cascade of system modules. The framework is extended to more general networking models for control, consensus, learning, and inference in  Sec.~\ref{sec:net}. Finally, we conclude the paper in Sec.~\ref{sec:concl}.

\section{The Context for Defining Timing Measures}\label{sec:definition}

In this section, we classify the notions of timing along multiple dimensions that depend on the context of the application or the communication service. 

\subsection{Time, Real-Time, and Simultaneity}
In its everyday use, the term \emph{real-time} is mostly associated with the sensation or perception of seeing things happen ``instantly'' or ``simultaneously''. For example, a sensor measuring our home temperature gives us the ``feeling'' of monitoring what is happening at home in real-time. The same applies to online chat applications, where we feel like talking in ``real-time'' with the other person. So what does \textit{real-time} really mean and, more importantly, can we give a universal definition? The above description provides some sort of vague or general yet operational definition. However, any attempt to formalize it into a universal definition would, if possible, pass through a universal or absolute definition of simultaneity and time perception. 
Before attempting to give our definition of real-time, we first discuss several misconceptions associated with this term. 

Real-time is often used interchangeably with the term ``\emph{live}''. However, real-time and live are not the same. Think of an event (signal) transmitted live from Mars using an electromagnetic wave. On Earth, we are going to see that event with a minimum delay of around $4$ minutes and a maximum delay of around $21$ minutes, depending on the actual distance between Earth and Mars. 
Real-time is often associated with {highly} stringent latency requirements and extreme performance. Suppose that a robot is required to move $100$~m under a ``real-time'' constraint of $50$~s. This says nothing about the speed of the robot, as long as the travel duration remains below $50$~s. This is because real-time is associated with a deadline, which is not necessarily stringent. In a refined definition of the term, real-time communication {simply} means that information or data (a message/packet or a set of messages) has to be transmitted and received on time, within a certain interval; not earlier, not later. 

Timing is related to communication latency, whose operational definition is the time required for a packet to arrive from its sender to the destination. Measuring this time difference implies the use of a common reference and clock synchronization, which is often not available in practice. On top of that, real-time brings the notion of deadlines and predetermined time instants into the picture. Therefore, a proper definition of timing requires an understanding of two important concepts: synchronicity and simultaneity, that is, the relation between two events assumed to be happening at the same time in a given reference frame. The former is relatively well understood in communication systems, and is often taken for granted. The latter is rather unexplored in wireless networking, and its relativity could bring new and interesting concepts and insights. Moreover, these two concepts bring up the theme of causality and space-time contiguity. Time at a particular location is defined by the measurement of a clock located in the immediate vicinity and is related to a certain reference frame. Every event that is spatially infinitely close to the clock can be assigned a time coordinate. Only the times of events occurring in the immediate vicinity of the clock can be ascertained directly by means of the clock. This means that at this moment one only has a notion of time in the vicinity of the chosen clock, which is one of the main observations in the Special Theory of Relativity \cite{SpecialRelativityBook}. 

\begin{figure}[t]
 \centering
 \includegraphics[width=\textwidth]{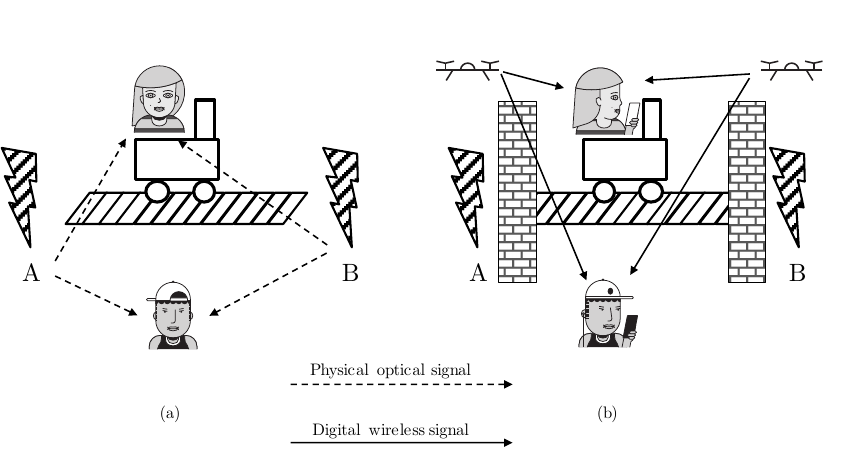}
 \caption{The relativity of simultaneity. (a) Communication through a  propagation of the physical optical signals. (b) Communication through a wireless digital system.}
 \label{fig:LightingTrain}
\end{figure}

Fig.~\ref{fig:LightingTrain}a illustrates the classic Einstein's example of the relativity of simultaneity. The static observer is at an equal distance from points $A$ and $B$. A lightning strikes each of those points and the static observer claims that the two strikes have occurred simultaneously. The moving observer sits on a fast train that moves towards $B$ and she claims that lighting has first struck $B$, and then another bolt struck $A$. The tacit assumption made by A. Einstein is that difference in the observations is solely due to the physical propagation of the optical signals that carry information about the lightning bolts. This means that both observers have identical instruments for registering the lightning and there is no difference in their observation due to, e.g. variations of the processing done in the measurement devices. 

In Fig.~\ref{fig:LightingTrain}b, the setup is changed. The spatial points $A$ and $B$ are shielded by tall walls, such that no visual information can arrive to the two observers. However, at each wall there is a drone that captures a video of the respective lighting and transmits the video through wireless connections to both observers. The digital receiver of each observer uses a certain playout delay to make the events video screen seem as if they occurred simultaneously.
Now both observers agree that the two events have occurred simultaneously, which is a digital distortion of the physical reality.   

This parallel with the Special Theory of Relativity indicates that simultaneity and causality, as well as its bi-directional relation with time, are  {essential} to defining timing. Drawing well-thought and operational analogies between timing in communication systems and time in physical systems (relativistic physics) and biological systems (horizon of simultaneity) could radically transform the notion of timing and synchronicity in future communication systems. This shift in thinking may {lead to the development of} a more general mathematical theory of timing in communications, one of the most difficult and important challenges remaining in communication theory.

\subsection{Timing Scales and Requirements}\label{ssec:realtime}
Timing requirements, expressed as, e.g., latency or jitter,  have traditionally been part of the set of \gls{qos} parameters defined for a given communication system, especially for applications tagged as real-time. 
However, as we discussed above, latency requirements and real-time constraints are highly dependent on the application, and different standards define different timing requirements. For example, the aim of 5G is to provide \gls{urllc} service for small data payloads (e.g. 32 bytes) with a maximum radio latency of $1$~ms (i.e., the latency is measured at layer 2 or 3) and reliability higher than 99.999\%. As wireless systems evolve beyond 5G towards a loosely defined set of technologies denoted as ``6G'', there is a general tendency towards supporting lower latency and operating at shorter, ms or sub-ms timing scales~\cite{adeogun2020towards}. 

\begin{figure}[t]
 \centering
 \input{./figures/oran}
 \caption{Timing scales of the \gls{ran} Intelligent Controller in O-RAN~\cite{oranrt}.}
 \label{fig:ORANRIC}
\end{figure}
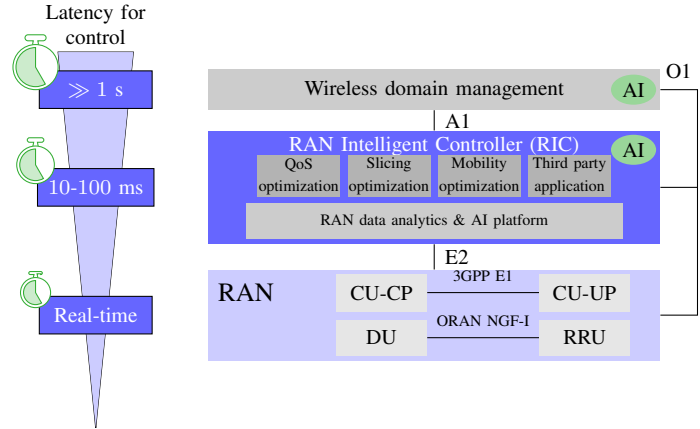

In order to define the relevant timing scale, we can follow the categorization used by the \gls{oran}~\cite{oranrt}, which defines three time-scale categories (see Fig.~\ref{fig:ORANRIC}): \emph{(i)}, real-time, \emph{(ii)}, near real-time, and \emph{(iii)}, non real-time. A similar classification is provided by the \gls{5gacia}~\cite{5gaciakey}, where the three categories are \emph{(i)}, hard real-time, \emph{(ii)}, soft real-time, and \emph{(iii)}, non real-time. Here we provide a slightly more general view on these timing categories:  
\begin{itemize}
    \item \emph{Real-time:} A universal definition of ``real-time'' is elusive, not to mention that it is often associated with speed and the notions of ``live'' and ``interactive''. Real-time does not necessarily mean that information can be exchanged instantly or with negligible latency. Although it may entail ultra-fast response time or immediate actions, its foundational element is that of completion in a predetermined, guaranteed amount of time. As such, real-time means controlled rather than zero latency. Real-time comes along with latency ``determinism'' and behavior predictability, which enable guarantees of achieving specific deadlines, being more or less stringent. For example, in the context of \gls{oran}, real-time denotes the processes (MAC scheduler or power control) for which the latency/timing measure is below $10$ ms, while in the context of the \gls{5gacia} requirements, hard real-time deals with timing on the order of ms or even \si{\micro\second}. 
    \item \emph{Near real-time}: This is also denoted as soft real-time, where the term ``soft'' denotes a relaxation in both the absolute timing horizon, allowing for longer latencies, and the level of determinism in the timing requirements, allowing for softened probabilistic guarantees. In terms of timing horizon, near real-time in \gls{oran} deals with timings between $10$ ms and $1$ second, while soft-real time in \gls{5gacia} can allow latencies on the order of a second. For instance, in \gls{oran}, near real-time may involve mobility or interference management. The real-time versus near real-time dichotomy can be interpreted as an effect of the cost of delayed action: if delaying an action is costly, the system should provide stricter guarantees, leading to harder real-time requirements. As near real-time backs away from almost deterministic latency guarantees, it also encompasses applications that are sensitive to the freshness of the data and \gls{aoi}. 
    \item \emph{Non real-time:} This refers to the case in which timing parameters are such that no latency or deadline guarantees can be provided. In both the \gls{oran} and the \gls{5gacia} definition, non-real time refers to timings longer than a second. Non real-time is associated with applications and procedures that are not time-sensitive and are denoted as best-effort~\cite{koutsiamanis2018best} or delay-tolerant~\cite{roy2018quality}. 
    \end{itemize}
    
Interestingly, the distinction between the above categories or the boundaries could be seen under the prism of effectiveness in achieving a specific goal. Timing requirements are usually imposed by services and may differ depending on the end user's perception or tolerance. Discrete automation and motion control may need end-to-end latencies of $1$--$5$~ms, whereas process automation (remote control, monitoring) could operate with $50$~ms latency. Specifically, according to ITU Network 2030 \cite{ITU_Network2030}, the upper bound on the end-to-end networking latency for haptic applications is on the order of $5$~ms or less. This allows for round-trip control loops that allow feedback-based haptic applications to operate under $10$~ms, even as low as $1$~ms in some cases. Autonomous mission-critical infrastructure relies on similar latency objectives. Industrial and robotic automation requires not only ``not-to-exceed'' latency, but an effectively ``deterministic'' latency, requiring predictability. This goes beyond in-time delivery; packets should be delivered ``on time'', i.e., not exceeding a certain latency but not arriving any sooner~\cite{bartolomeu2018supporting}. Industrial automation systems (Industry X.0) are based on real-time enabled \glspl{cps}, which will serve as platforms connecting people, objects, and systems to each other. Latency requirements for different applications range from several ms for mechanics, to several ms down to $1$~ms for \gls{m2m}, to 1 ms for electrics \cite{AccReport}. In \gls{v2v} networks, the time needed for collision avoidance in safety applications is below $10$~ms \cite{simsek20165g}. In case a bidirectional data exchange for autonomous driving maneuvers is considered, a latency on the order of $1$ ms is most likely needed. In \gls{v2x}, messages for situational awareness, e.g., \gls{cam} and \gls{bsm}, are generated periodically (commonly every $100$~ms) including vehicle state information such as geolocation, velocity, heading and other related information.
In e-healthcare applications, an end-to-end latency of a few milliseconds, together with ultra-high reliability in wireless link connection and data transmission is required. In online gaming, latency around $20$ to $100$~ms could still provide satisfactory gameplay experience, although lower latency is needed for maximum performance in games where timing is important. 
The latency requirement of holographic communications is on the order of $10$ ms to allow instant viewer position adaptation at $60$~\gls{fps}. However, the latency requirement can be relaxed, becoming as low as conventional interactive video (on the order of $100$~ms).

An example of timing-oriented networking design is \gls{tsn}, poised to connect and transform today's factories~\cite{larranaga2020analysis}. 
\gls{tsn} refers to a group of networking protocols and standards developed by the IEEE 802.1 \gls{tsn} working group to provide accurate time synchronization, hard real-time constraints, and zero congestion loss in \glspl{lan}. \gls{tsn} handles three main functions: synchronizing all the clocks on the network, scheduling the most important traffic, and “shaping” the remaining traffic to achieve the desired traffic patterns. Taking \gls{tsn} standards, which have been developed mainly assuming Ethernet as the underlying communication medium, the 3GPP has made significant progress in the last releases to complete the integration with 5G~\cite{3GPPTS23.501}. %
A limitation of \gls{tsn} is that deterministic service is provided over a short distance. Moreover, \gls{tsn} is geared towards \gls{cbr} traffic, not \gls{vbr} traffic.

\subsection{Timing and Communication Actors}

Through the description of the timing scales and requirements it becomes apparent that communication actors represent an important factor that determines the perception of timing in a communication system. 
For example, real-time for machines that have sub-ms reaction times~\cite{houska2011auto} has a different meaning than real-time for systems with a human in the loop, where latency longer than $50$ ms could be tolerated. Then, is there a universal or optimal value for latency and reaction time? The answer depends on the context and the communication actors (human or machines). Note that the term ``machine'' should be understood in a broader sense, beyond that of a simple man-made, electromechanical device. As such, a program or software application can also be treated as a machine in this context.

Depending on the actors and the communication parties involved, we can have the following first-order classification: 
\begin{itemize}
    \item \emph{Human--Human:} 
    In scenarios where humans communicate and interact with other humans, the timing and reaction time limits depend on the characteristics and the limitations of human senses, as well as on humans capabilities in terms of sensory perception, cognition, and physical and neural transmission and processing times. For example, the neural processing time differs between the senses, and it is typically slower for visual stimuli than for auditory ones; approximately $50$~ms and $10$~ms, respectively. For touch, the brain may have to take into account where the stimulation originated, e.g., toes, nose, etc., as traveling time to the brain is not the same. Our brain can only process an image if our eye sees it for at least $13$~ms \cite{boccolini2019ghost}, which corresponds to about $75$~\gls{fps}, and receiving a stream of data faster than this will only underscore the limits of our perception. 
    Accordingly, the definition of \emph{real-time} for human communications has a hard limit given by perception: after video communications reach the perception threshold and achieve a reliable $10$~ms latency at 75~\gls{fps}, any further improvement in the communication system will not provide any benefits to the user in terms of \gls{qoe}.
    Providing exact values on this matter goes beyond the scope of this paper and is an ongoing research topic. Nevertheless, an intriguing and surprising aspect is that despite naturally occurring time lags and asynchronous arrivals of auditory and visual information, humans perceive inter-sensory synchronicity for most multi-sensory events in the external world, and not only for those within the so-called ``horizon of simultaneity'', i.e., a distance of approximately $10$ to $15$~m from the observer~\cite{Poppel88}.
    \item \emph{Human--Machine:}
    This scenario entails communication and interaction between humans and machines. Machines are expected to be ``faster'' than humans, which will then define the timing requirements, as the human perceptual system is the bottleneck of the system. An interesting aspect here is how time is perceived by humans when they are interacting with a machine. Various studies on human-machine interaction, starting from R. B. Miller's seminal work in 1968~\cite{miller1968response}, have shown that the average human reaction time is on the order of $250$~ms. Moreover, humans perceive a response time of $100$~ms as instantaneous, whereas uninterrupted flow\footnote{The definition of the term ``flow'' corresponds to \emph{``a state of concentration so focused that it amounts to absolute absorption in an activity''} \cite{Flow_book}. When we experience flow, we lose track of time, and time feels accelerated.} is experienced with a $1$~s response time. 
    \item \emph{Machine--Machine:}
    In this setting of increasing importance, machines are interacting with each other without the possibility of human intervention, and \gls{m2m} traffic is becoming an important class in mobile networks. As such, the timing requirements will exclusively be dictated by the limits of the specific machines. The absolute performance limits of machines are not fully known or understood, but machines are in general subject to the theoretical limits described by computational complexity theory and the laws of physics. 
\end{itemize}

Presently, there is a consensus that future communication networks will have to pass from human speeds to machine speeds; this will be even more emphasized as we are moving towards 6G communication systems \cite{SOTE_Report2020}. The distinction between real-time and non real-time optimization is also crucial for intelligent network optimization, as designing network elements that can cooperate distributedly and on different timescales is a complex task~\cite{banchs2021network}. 
The Internet as we know it and current wireless networks have been designed for humans: humans browsing web pages, exchanging emails and messages, watching movies, etc. Therein, we know that humans have limitations in terms of the visible spectrum (from $380$ to $780$~nm), the perceivable frame rate and resolution, and the audible frequency range (from about $20$~Hz to $20$~kHz).
``\emph{This is why today's Internet — while fast enough for most humans - appears glacial when machines talk to machines}''~\cite{SOTE_Report2020}. For example, an autonomous vehicle or a drone moving at $90$~km/h will travel $100$~m in $4$~s. Avoiding collisions may require ultra-fast decision-making: a delay of $100$~ms could cause it to crash into something as far as $3$~m away.
However, what are the limitations of machines in the context of wireless communication systems? What communicating and performing decisions and actions at machine speeds imply for the supported applications and services? 
We also note that the data generation process can vary significantly across communication actors. Some actors could generate ``small and bursty'' data, e.g., indicating a machine's status, whereas other actors or ``things'' (e.g., surveillance cameras) could generate very large amounts of data.

In addition to the involved communication actors, another classification considers who triggers the communication process, such that there are event-triggered and time-triggered systems, respectively. 
In event-triggered (real-time) systems, a processing or a transmission activity is initiated as a consequence of the occurrence of a significant event. An example of an event triggered system is an alarm system. In a time-triggered system, the activities are initiated periodically at predetermined points in time. An example of a time-triggered system is a production system with a pre-planned production cycle or a traffic light system that follows a strict timing schedule.
Event-triggered systems excel in flexibility, whereas time-triggered systems excel in temporal predictability. In event-triggered systems, the communication delay may be time-varying and quite susceptible to jitter. In time-triggered systems, it is essential to synchronize the actions of all participating nodes to a global time. Since the (off-line) scheduling predefines the time windows for all actions, the result is a time scheme with constant latencies and no jitter. If no synchronization is implemented, the latency and the jitter will most likely be of higher magnitude than for event-triggered systems.

\section{A Statistical Framework of Timing}\label{sec:framework}

An important element in defining a model for timing is the reference with respect to which timing is measured.  
In this section, we define the statistical framework for timing for the case of a single link, or even a multihop connection, between
Node 1 and Node 2. In order to keep things simple at this stage, we also assume that the clocks of Node 1 and Node 2 are perfectly synchronized, such that we can talk in terms of absolute time, as observed identically by both nodes. 

\subsection{Timing References and the Role of an Initiator}

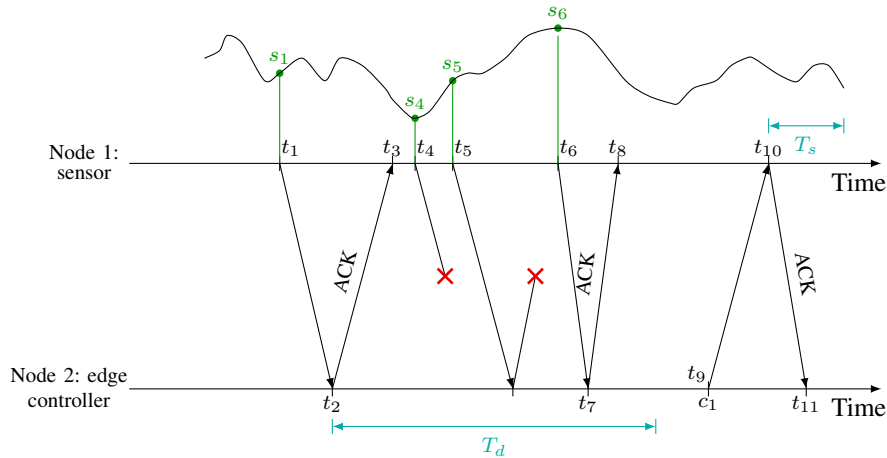
\begin{figure}[t]
 \centering
 \input{figures/references}
 \caption{Illustration of the timing references for a link between a sensor (Node 1) and edge controller (Node 2).}
 \label{fig:TimingRefsIllustration}
\end{figure}

Consider the simple communication scenario in Fig.~\ref{fig:TimingRefsIllustration}, in which Node 1 is a sensor that monitors a physical process and Node 2 is an edge controller. It is assumed that both nodes are synchronized and measure the time {identically}. Node 1 samples the physical process and sends updates to the edge controller. The sample $s_1$ is taken at time $t_1$, received by Node 2 at $t_2$ and acknowledged to the Node 1 at $t_3$. Node 2 is interested to have an update on the state of the physical process that is as fresh as possible, i.e., to minimize the \gls{aoi} with respect to the physical process observed by Node 1. When Node 2 receives $s_1$, its age is already $\Delta t=t_2-t_1$. Hence, Node 1 measures the age with respect to a past timing reference $t_1$, associated with the value of the process state. 

The system is programmed to work such that if the controller does not receive any packets from the sensor within a time interval $\Delta t=T_d$, it initiates a safety shutdown of the system. For the example on Fig.~\ref{fig:TimingRefsIllustration}, at time $t_3$ Node 1 learns that it must deliver at least one data sample to Node 2 before the deadline $t_2+T_d$, or the system will shut down. Due to transmission errors, $s_4$ is not received by Node 2. The sample $s_5$ is received by Node 2, but its acknowledgement is not received by Node 1, such that after $t_5$ Node 1 still considers the deadline to be $t_2+T_d$ and invests extra communication resources to deliver the data sample $s_6$, whose reception at time $t_7$ is acknowledged at time $t_8$.

Finally, at time $t_9$, the edge controller sends the command $c_1$ to Node 1 to go into sleep mode for an amount of time $\Delta t=T_s$ after receiving the command. Node 1 sends an acknowledgement and goes into sleep mode. 

For all communication instances from Fig.~\ref{fig:TimingRefsIllustration},  there is a certain time interval $\Delta t$ during which communication takes place. We will refer to it as a \emph{communication interval} and an important aspect is the timing reference with respect to which this interval is measured. The example illustrates three types of timing references:
\begin{itemize}
    \item \emph{Past Timing Reference}, or shortly, \emph{timing anchor.} This is the case when time is measured with respect to an instant that occurs in the past, such as the state of a monitored physical system. For example, \gls{aoi} is defined with respect to the timing anchor, as at the destination the anchor is the generation time of the last received update.
    \item \emph{Future Timing Reference} or shortly, \emph{deadline}. In this case the timing reference is at a point $t=T$ in future and it represents a certain deadline. The communication interval $\Delta t$ is then measured backwards, starting from the future moment. This reflects the fact that communication should start before time $T-\Delta t$ in order to meet the deadline.
    \item \emph{Relative Timing Reference.} In this case one or more of the nodes participating in the communication process can choose the reference moment $t=0$ and measure the interval $\Delta t$ relative to that moment. This is the example from Fig.~\ref{fig:TimingRefsIllustration} with the sleep command. 
\end{itemize}
For consistency, all these timing references are defined from a perspective of an external genie that can perfectly observe the system. In reality, the nodes can have discrepancy in their timing references and communication is used as a means to reconcile this discrepancy. For example, if Node 1 decides to denote a certain time as $t=0$, then Node 2 does not know this until it receives a packet from Node 1. From the way they are defined, the past and the future timing anchors have a direct relation to a time instant in the physical world and are related to a sensing/actuating operation through which the digital system interacts with the physical world. Differently from this, the relative timing reference is mostly related to the ``digital time'', as measured by the digital systems of the involved devices. For instance, the go-to-sleep command is not related to an event in the physical world, but it is initiated by a digital command conceived in the edge node.

Another important question is that of who plays the role of the \emph{initiator} of the communication. For example, when Node 1 reports the status of a physical process, it is Node 1 that initiates the process. In a different case, if Node 2 sends a query that demands some information from Node 1, then the initiator is the receiver (Node 2). Depending on who has the role of the initiator, there are, in general, two types of communication: 
\begin{itemize}
    \item \emph{Push-based communication}, where the initiator is the information sender.
    \item \emph{Pull-based communication}, where the initiator is the information recipient. 
\end{itemize}
At a first glance, push-based communication can be associated with a timing anchor or can be triggered by an event, while pull-based communication with a future deadline. However, this is not necessarily the case. For example, think of the case in which Node 1 is a controller that wants to put Node 2 in a certain state at a future instant $T$. This is a push-based communication with a future deadline. As another example, Node 2 can send a query to ask for the most recent state of the system: this is a pull-based communication with a past anchor.

\subsection{Statistical Characterization of Timing Measures}

Let us assume that Node 1 observes the physical system at time $t=0$, creates a packet of size $D$ bits, and transmits it to Node 2. 
The communication interval $\Delta t$ starts at $t=0$, and it is convenient to describe the stochastic behavior of the connection by a \emph{latency-reliability} function  
\begin{equation} \label{eq:LatencyReliability}
    F_D(\tau)=\Pr(\Delta t \leq \tau \mid D),
\end{equation}
which is a non-decreasing function that denotes the probability that the packet of size $D$ is received and processed correctly at Node 2 by the time $t=\tau$. Intuitively, this function reflects the fact that, as time passes, Node 1 has more actions at its disposal to increase the probability that the packet is decoded by Node 2. If packets are never dropped and are always delivered without errors, i.e., $\lim_{\tau\to\infty}F_D(\tau)=1$, the latency-reliability function is equivalent to the concept of a statistical delay bound, which is widely used in stochastic network calculus~\cite{Fidler2014NetworkCalculus} to analyze delay violation probabilities in multi-hop queuing networks with stochastic arrival and service times.%

The above stochastic model can be generalized by considering a more complex event in the communication system rather than reception of a single packet. For example, in a multicast scenario one can look at the time interval in which at least $K$ nodes have received a certain data packet. Similarly, if there is a transmission of a batch of files, the event of interest can be the reception of at least $L$ files from the batch. An interesting scenario is when reconstruction requires a specific ordered sequence of $L$ packets carrying correlated information. Therein, timing measures have to be revisited; if packets do not arrive consecutively, timing (\gls{aoi}) is measured as the difference between the current time and the generation time of the latest ``entirely'' received correlated sequence of packets.  Further generalization can be made by considering a \emph{prior context} ${\cal C}$ of the system instead of only a packet of size $D$. An example of a context is a prior knowledge that a node may have. Another example is the context in which Node 1 has the data file $\mathbf{D}_1$ and Node 2 has the data file $\mathbf{D}_2$, and we are looking at occurrence of the event in which both nodes have both files. The event we are looking at will be clear from the prior context, such that we can write
\begin{equation} \label{eq:TimingContext}
    F_{\cal C}(\tau)=\Pr(\Delta t \leq \tau \mid {\cal C}),
\end{equation}
which, like \eqref{eq:LatencyReliability}, is a non-decreasing function.

In order to expand the set of relevant statistical measures, recall that two basic problem categories in statistical modeling are \emph{statistical decision} and \emph{statistical estimation}, respectively. In the context of timing in communication systems, the above discussion is limited to discrete events and statistical decisions and finding the probability that some event has taken place. A completely different set of problems is obtained when we put the statistical estimation in the context of timing. 

To illustrate this, let us take a timing anchor. At time $t$, Node 1 measures a certain state, registers the value $x(t)$ and communicates the state to Node 2. The estimate that Node 2 has about the state $x(t)$ of the physical system after a communication interval of $\Delta t$, is denoted by $\hat{x}_{\Delta t}(t)$. The quality of this estimate after the communication interval $\Delta t=\tau$  can be measured by a generic loss function $L(\hat{x}_{\tau}(t),x(t))$, which should increase over time, at least in the average sense. The \gls{mse}, i.e., $\|\hat{x}_{\tau}(t)-x(t)\|^2$, is a common way of measuring this estimation error, but other loss functions can be used.

To support this observation, one can think of a communication strategy that continuously sends refinements from Node 1 to Node 2 about the state observed at a past anchor $t$. Alternatively, consider the special case in which Node 1 creates a single packet to describe $x(t)$ and this packet is an atomic unit of communication. In this case, $L(\hat{x}_{\tau}(t),x(t))$ has a particular form: it has a positive value (e.g. based on a prior knowledge that Node 2 has about $x(t)$) until $\tau=\tau_0$ that corresponds to the time $t+\tau_0$ at which Node 2 receives successfully the packet from Node 1. For $\tau > \tau_0$ it is $L(\hat{x}_{\tau}(t),x(t))=0$ or, possibly the quantization error for $x(t)$. As another example, in a setup with distributed learning, the true $x(t)$ is not known to any of the nodes, but the (empirical) loss decreases as learning progresses in time. In the opposite case, in which the state is high-dimensional (e.g., an image or depth map of the environment in a cooperative driving scenario) or the observation is distributed among different sensors, even the definition of the loss can become complex, and decisions need to be made based on which piece of information is more important at a given moment, i.e., which transmission results in the biggest reduction in the loss function, which never goes to zero. Finally, with respect to timing relativity and simultaneity, in remote actuation and distributed real-time systems, we need to minimize $L(\hat{x}_{\tau}(t),x(t))$ for small $\Delta t+\mathcal{T}$, where $\mathcal{T}$ could include time spent for information generation, processing, and reconstruction \cite{kountouris2020semantics}.

\subsection{Summary of the Basic Framework}
Our framework for describing the timing problems in communication systems will rely on the timing reference and the statistical operation (decision or estimation). In order to keep the discussion compact, we do not use the role of initiator to add a third dimension, but we will use it as a supplementary information where relevant. 

\begin{itemize}
    \item \emph{Timing anchor.} 
    \begin{itemize}
        \item \emph{Statistical Decision.} Node 1 sends updates to Node 2 about the state of a monitored physical process. A relevant timing measure is \gls{aoi}. This can be push-based, such that Node 1 decides when to send an update and attempt to ensure that Node 2 always has the freshest update on the status of the process. Alternatively, it can be pull-based, such that Node 2 sends queries to demand status updates.
        \item \emph{Statistical Estimation.} Consider a case similar to the previous one, where Node 2 receives updates from Node 1 about the state of a certain physical system. However, the state at time $t$ is a multidimensional variable and cannot be accommodated in a single packet transmission, but rather sent gradually. Hence the correctness of the estimate that Node 2 has about the state at time $t$ will increase over time. In a push-based communication, Node 1 initiates the transmission and transmits either until receiving a stop feedback from Node 2 or until estimating that Node 2's estimate about the physical system is sufficiently correct. In the pull-based case, Node 2 initiates the communication and, as it receives data from Node 1, it judges the quality of the estimate and, if it is not satisfactory, sends further pull requests to require more data. 
    \end{itemize}
    \item \emph{Timing deadline.} 
    \begin{itemize}
        \item \emph{Statistical Decision.} This is the classical case of a latency constraint, where a data packet should be delivered within a given deadline. The timing requirements of URLLC are defined in this context, as the packet is considered to be ready for transmission and needs to be delivered within a deadline (e.g., $1$~ms).
        \item \emph{Statistical Estimation.} Here the receiver wants to estimate a certain variable within a given deadline and with error no larger than a certain $\epsilon$. One example from satellite communication entails a satellite that is visible for a limited time period, as the estimation needs to have acceptable accuracy until the link becomes unavailable. 
    \end{itemize}
    \item \emph{Relative Timing Reference.} 
    \begin{itemize}
        \item \emph{Statistical Decision.} This is the case in which a group of nodes want to reach a consensus on a decision and the set of possible decisions is discrete. For instance, the decision could be related to the precedence among the autonomous vehicles at a traffic crossing or to which blockchain transaction is considered valid. 
        \item \emph{Statistical Estimation.} A use case that falls into that category is distributed learning. Therein, the model training among nodes should be completed within a given interval from the time the first node has initiated the process, where completion is declared based on a certain threshold on the measure of loss.  
    \end{itemize}
\end{itemize}

Another level of complexity is revealed when we start to ask other questions: what does one node know about the knowledge of another node? In the case with a past anchor, Node 1 observes the state of a physical system $x(t)$ and sends it to Node 2, which in turn makes an estimate $\hat{x}_{\tau}(t)$. One related question is: what does Node 1 know about the value of $\hat{x}_{\tau}(t)$?
In a simple case, if Node 2 receives the packet successfully from Node 1 after an interval $\tau_1$ and sends an ACK that requires time $\tau_2$, then Node 1 knows $\hat{x}_{\tau_1+\tau_2}(t)$ perfectly. This is important in, for example, status monitoring application where Node 2 needs to take an action based on the current state of Node 1. Then, Node 1 may know what the status is only after time $t+\tau_1+\tau_2$. If Node 2 cannot decode the message and sends instead a NACK, then Node 1 knows the last correctly received status $\hat{x}_{\tau_1+\tau_2}(t-\Delta)$, transmitted from Node 1 to Node 2 at $t-\Delta$. Note that, upon transmission failure, Node 1 has the option to resend the same data and thus potentially use some combining with the previously received version of the data. Alternatively, retransmissions of the same data are dropped and, upon failure, the status of the process monitored by Node 1 is sampled anew and transmitted. Two-way communication is further discussed in Section~\ref{sec:two-way}.

\section{Putting the Prior Art within the Statistical Framework}\label{sec:sota}

Now that we have defined the basic framework of timing measures, we can look at the existing body of work on timing in communication networks, trying to frame the rich literature into the categories defined in the previous section.  Some recent standards, including those developed by \gls{3gpp}, are beginning to consider these factors and metrics in a more general way, as shown in the diagram in Fig.~\ref{fig:timing_model}, which is adapted from \gls{3gpp}. The model includes some of the concepts that we will discuss in the later sections, such as the notion of timing at the higher and lower layers, as well as the importance of the inter-transmission time, defined as ``transfer interval'' by \gls{3gpp}. In the following, we examine a few interesting cases, which are familiar to the networking community and include the notions used by \gls{3gpp}. Our framework can subsume all these metrics in the same perspective and allows us to think holistically about timing and the related metrics. The relevant references are summarized by topic in Table~\ref{tab:sota}: as the table shows, past anchors are the most common method of measuring timing and are often used in standards and protocols, while the use of relative timing references, which consider complex networking scenarios, is still largely unexplored.

\begin{table}[t]\centering
  \renewcommand{\arraystretch}{1.3}
  \caption{Representative works grouped by their relation to the instances of the statistical framework.}
  \label{tab:sota}
    \footnotesize\centering
	\begin{tabular}[c]{lll}
		\toprule
		Topic & Timing reference & Significant references \\
		\midrule
		Latency in queuing systems & Past anchor (stat. decision) & \cite{briscoe2014reducing,DimitriouTWC2018, parvez2018survey,Nain1985analysis,Behroozi1992delay,NawareTIT2005,GeorgiadisJSAC1987}\\
		Statistical latency guarantees & Past anchor (stat. decision) & \cite{Chang1994stability,Wu2003effectivecapacity,Soret2010correlated,Soret2009voice}\\
		End-to-end latency in realistic systems & Past anchor (stat. decision) & \cite{PappasTWC2015,Stamatiou2013delay, Fidler2014NetworkCalculus,polese2019survey}\\
		\midrule
		Latency deadlines & Deadline (stat. decision) & \cite{BhattacharyaTAC89, HouINFOCOM2009, hou2013packets,popovski20185g}\\
		Timely throughput & Deadline (stat. estimation) & \cite{LashgariTIT13,YangMobihoc2019,Soret2014fundamental}\\
		Deadline-based optimization &  Deadline (stat. decision) & \cite{AnandJSAC2018,FuTWC2006,EwaishaTVT2017,FountoulakisGC2018,NeelyTAC2013,TsanikidisINFOCOM20, pocovi2017mac, Soret2018ICIC}\\
		End-to-end deadlines & Deadline (stat. decision) & \cite{FountoulakisWiOpt2017,chiariotti2019analysis,chiariotti2021hop,guo2017dems,bentaleb2018survey}\\
		\midrule
		\gls{aoi} & Past anchor (stat. decision) & \cite{kosta2017age,maatouk2020age,yates2020age,chen2020age,KostaTCOM20,SunJCN19,yates2020agenet,zhou2020risk,devassy2019reliable,chiariotti2021peak, KostaJSAC21, FountoulakisITW20,yates2020agesurvey}\\
        Goal-oriented \gls{aoi} extensions & Past anchor (stat. estimation) & \cite{maatouk2020AoII,ayan2019age,zheng2020urgency}\\
        Pull-based \gls{aoi} & Past anchor (stat. estimation) & \cite{holm2021freshness,li2020waiting}\\
        \midrule
        \gls{3gpp} service requirements & Deadline (stat. decision) & \cite{popovski20185g,3GPPTR38.913,3GPPTS22.261,3GPPTS22.104,miremadi1995evaluating}\\
        \gls{ttff} & Past anchor (stat. estimation) & \cite{kruczynski1995time,paonni2010performance}\\
        Synchronization requirements & Past anchor (stat. estimation) & \cite{levesque2016sync,3GPPTS23.501,5gaciakey,5gaciatsn,rfc5905,rfc3550}\\
        \midrule        
        Distributed learning requirements & Deadline (stat. decision) & \cite{GX:AirAgg:2020,Quek:ScheduleFL:2020,Zeng:EnergyEfficientFL:ArXiv,Ren:ImportantAwareFL:2020}\\
        Distributed learning speed & Relative (stat. estimation) & \cite{bernstein2018signsgd,Gx:OneBitAirAgg:2021}\\
		\bottomrule
	\end{tabular}
\end{table}

\subsection{Latency}
Latency, also known as delay, is perhaps the simplest and oldest metric used to measure timing in networks; it captures the time that a packet spends in the network. Latency is measured with respect to an event that happened in the past and it characterizes packets. 
In our framework, it is an example of a past timing reference: the latency timer starts when the packet is transmitted by a given layer in the protocol stack and stops when the packet arrives to the same layer at the destination. A closely related metric is the \gls{rtt}, which represents the latency over both sides of the connection, from the moment the packet is sent to when the transmitter receives the related acknowledgment. Latency has been studied extensively in different system setups, both theoretical and practical. An exhaustive literature review on the topic is outside the scope of this work, but we will list some relevant works and refer the reader to existing surveys for a deeper examination~\cite{briscoe2014reducing,parvez2018survey}.

The main theoretical tool for analyzing latency in networks is queuing theory, which can go from simple $M/M/1$ systems~\cite{Nain1985analysis} to complex access mechanisms with different arrival patterns~\cite{Behroozi1992delay}. In particular, random access mechanisms such as ALOHA~\cite{NawareTIT2005}
have been extensively studied~\cite{GeorgiadisJSAC1987}. One of the first additions to plain communication latency was the observation that applications like video are also sensitive to the jitter, defined as the variation in latency of the packet flow~\cite{balakrishnan1962problem}. Another extension is to derive bounds on the tail of the latency distributions, which can provide statistical \gls{qos} guarantees, such as effective bandwidth/capacity and bounds in queue length and latency violation probability~\cite{Chang1994stability, Wu2003effectivecapacity}. The analyses get complex when the intricate correlations in the arrival and/or the channel processes are considered~\cite{Soret2010correlated, Soret2009voice}.

It is also possible to study end-to-end latency, going beyond a single link and looking at the connection level. In this case, there are fewer theoretical works analyzing the latency under realistic access networks; they mostly consider the two-hop case~\cite{PappasTWC2015} or Poisson traffic~\cite{Stamatiou2013delay} due to the complexity of analyzing latency in other scenarios. Even small random access networks with bursty traffic become rather intractable due to the coupling among the queues \cite{DimitriouTWC2018}, a scenario that still remains largely unexplored. An alternative to address this complexity is using stochastic network calculus~\cite{Fidler2014NetworkCalculus}, which is a probabilistic extension of network calculus~\cite{LeBoudec2001networkcalculus,cruz1991networkcalculus,cruz1991networkcalculusII}. Network calculus builds upon dioid and $(\min,+)$ algebra and provides backlog and delay bounds to understand the statistical multiplexing and scheduling of non-trivial traffic sources. Its stochastic counterpart has been extensively employed to analyze in wireless networks in various settings with time-varying random service rate~\cite{SNC1,SNC2,SNC3}.

Several learning-based mechanisms to reduce latency have been proposed for 6G networks, often including computation as well: the placement of computation tasks and network functions is critical for reducing latency in complex tasks~\cite{kaur2022latency} and the joint consideration of computational and communication aspects can lead to a lower overall latency for different services~\cite{liao2020distributed}. Domains that have strict constraints and high throughput, such as \gls{vr} transmission~\cite{fantacci2021end}, or fast mobility, such as vehicular communications~\cite{zhou2021drl}, pose specific challenges that need to be addressed individually. The use of digital twin models~\cite{sun2020reducing} of the network can also improve the effectiveness of learning-based optimization schemes, providing more training data and reducing the impact of the training on the real network. These models can be used with any kind of learning models, such as federated learning, and improve resource allocation for complex distributed applications such as \gls{dlt}~\cite{lu2020low}.

The latency is minimized by reducing the time that a packet spends in the network, including the initial access delay, which can be large in wireless networks. In the queuing models mentioned above, this is equivalent to decreasing the total time spent in queues throughout in the network, which can be done by increasing the service rate and decreasing the rate at which packets enter the network. However, reducing the latency comes at the cost of reducing the throughput of the network, and finding the optimal trade-off is a central topic in many of the works above. At a more practical level, minimizing the end-to-end latency has also been one of the main goals of the recent research on transport protocols. Congestion control mechanisms are often too aggressive and overshoot the available capacity, causing significant increase in the latency -- see~\cite{polese2019survey} and the references therein. The practical role of congestion control in terms of latency is in the shaping of the traffic, which in turn affects the state of the queue and, consequently, future decisions from all transmitters. This tight coupling makes the use of a metric as old and traditional as latency an interesting yet challenging research avenue. 

\subsection{Deadline-constrained traffic}
The rise of real-time or near real-time and interactive applications has revealed the fact that minimizing latency is not sufficient for the smooth operation of such applications. Specifically, the network should operate with respect to \emph{deadlines}~\cite{BhattacharyaTAC89, HouINFOCOM2009, hou2013packets}. Deadline-constrained traffic is an example of a future timing reference. When a packet is generated by the transmitter, the timer does not move forward, but backward from the maximum allowed latency $T$. If a packet arrives within the deadline, i.e., before timer reaches zero, the transmission is successful, and thus the latency of the packet is irrelevant as long as it arrives within the deadline. The \gls{urllc} traffic class~\cite{popovski20185g} in 5G and beyond systems is a classic example of deadline-constrained traffic, which is relevant in industrial scenarios.

A metric called timely throughput~\cite{HouINFOCOM2009,LashgariTIT13} measures the amount of traffic that can be successfully delivered within the deadline, potentially including the effects of computation~\cite{YangMobihoc2019}. In general, there is often a trade-off between the achievable throughput and the tightness of the deadline $T$, as setting a tighter deadline requires more resources for every single packet~\cite{Soret2014fundamental}. Naturally, the achievable deadline has a hard floor given by the minimum latency in perfect conditions: while 5G and beyond systems envisage to achieve deadlines below 1~ms, the technical challenges may involve computational components and hardware limitations, as well as the effect of the medium access mechanism used.

As the timing reference is in the future, it is not necessarily ideal to handle the packets according to a \gls{fcfs} policy. As a result, optimizing the communication for deadline-constrained traffic is more involved than simply minimizing the latency. Most works that deal with deadlines aim at optimizing the medium access\cite{pocovi2017mac}, resource provisioning\cite{AnandJSAC2018}, interference management~\cite{Soret2018ICIC} and packet scheduling \cite{Destounis18} to reduce the deadline violation probability~\cite{FuTWC2006}. Another example is the optimization of error targets in deadline-constrained \gls{harq} protocols for URLLC~\cite{kotaba19}. It is also possible to jointly optimize the scheduling with other transmission parameters, such as power control~\cite{EwaishaTVT2017,FountoulakisGC2018}. More advanced schemes include the use of \glspl{mdp}~\cite{NeelyTAC2013} and randomization~\cite{TsanikidisINFOCOM20}, combining the scheduled approach with adaptive techniques that alter scheduling decisions to fit traffic patterns.

It is also possible to impose deadlines on end-to-end traffic, providing probabilistic guarantees or adapting the sending rate to make sure that packets meet the deadline~\cite{FountoulakisWiOpt2017, chiariotti2019analysis,chiariotti2021hop}. In this case, multiple connections are often used along with packet-level coding~\cite{guo2017dems}, considering the latency not in terms of a single packet but of an application block. If we go even higher on the protocol stack, an interesting case is given by HTTP Adaptive Streaming (HAS)~\cite{bentaleb2018survey}, a video streaming protocol at the application layer. Therein, the deadline is not fixed, as it does not represent interactivity, but depends on the state of the video playout buffer at the receiver: in order for the video to play smoothly, the transmission must be completed before the available video segments finish playing. This is an example of a relative timing reference, as the deadline for each block of data depends on the content of the packets themselves and on the state of the playout buffer.

\subsection{Age of Information}\label{sec:aoi}
Research on \gls{aoi} has seen a remarkable development over the past decade~\cite{kosta2017age}. In its original form, \gls{aoi} refers to the timing metric that describes the \emph{age} of the most recently received packet at the destination. Specifically, consider a sequence a packets generated by the source at times $\tau_1,\tau_2,\ldots$, and received by the destination at $\tau_1',\tau_2',\ldots$, respectively, and denote the generation time of the most recent packet by
\begin{equation}
\xi(t)=\max\{\tau_n \mid \tau_n'\le t\}.
\end{equation}
The \gls{aoi} at time $t$, usually denoted $\Delta(t)$, is then defined as
\begin{equation}
\Delta(t)=t-\xi(t),\quad t\ge \tau_1'.
\end{equation}
Since packet generation times and transmission latency are usually random, $\Delta(t)$ is a random process with sample paths that increase linearly between packet receptions, leading to the characteristic sawtooth pattern often associated with \gls{aoi}.

The \gls{aoi} metric fits well into our framework by defining the timing anchor to be the instant at which the most recently received packet is generated, i.e., $\xi(t)$. Note that the timing anchor is updated every time a new packet is received. This is different from the traditional latency metric where the anchor is updated when the transmitter generates a new packet.

The fact that \gls{aoi} measures the freshness of the information available at the receiver makes it a representative metric for remote monitoring and control-oriented tasks. The sawtooth pattern of linear \gls{aoi} has led to many works deriving its average in different systems, such as \gls{csma}~\cite{maatouk2020age}, ALOHA~\cite{yates2020age}, and slotted ALOHA~\cite{chen2020age} networks. Furthermore, the notion of \gls{aoi} can be generalized to measure any non-decreasing function of the age~\cite{KostaTCOM20, SunJCN19}. More specifically, as discussed in \cite{KostaTCOM20}, non-linear aging functions can implicitly capture the autocorrelation structure of the source signal. That work considers and analyzes three cases, the linear, the exponential, and the logarithmic as aging functions. When the autocorrelation is small, the exponential function can be a relevant choice, since it penalizes the increase of system time between updates, which in turn will affect significantly the remote reconstruction of that process. If the autocorrelation is large, then the logarithmic function becomes more relevant, whereas for intermediate values, the linear case can be a reasonable choice. This is also a step towards the effective age and the importance of information \cite{Agheli2022}, discussed in detail below.
\gls{aoi} has also generated several related metrics, which usually consider relative timing references. The most common example is \gls{paoi}, which samples the \gls{aoi} immediately prior to the reception of a new packet. Using the notation introduced above, the \gls{paoi} is the \emph{discrete} random process $\Delta'(\tau_2'), \Delta'(\tau_3'),\ldots$, where $\Delta'(\tau)=\lim_{t\to\tau^-}\Delta(t)$.
This is useful when measuring the worst-case performance of a system, and particularly, due to its analytical tractability, when considering not just the mean, but higher moments of the age distribution~\cite{yates2020agenet}, metrics related to its tail~\cite{zhou2020risk}, or even the complete \gls{pdf}~\cite{devassy2019reliable,chiariotti2021peak, KostaJSAC21, FountoulakisITW20}.
 \gls{paoi} captures the key characteristics of the age process. Furthermore, as shown in \cite{chen2021OJCOMS, KostaJSAC21}, the average \gls{paoi} and the average \gls{aoi} coincide in discrete time systems under the generate at will policy of status updates.

Gossip is a mechanism to convey information, such as status updates, in distributed systems and networks. Thus, considering timeliness and freshness metrics in such setups becomes relevant and important. The work in \cite{ioannidis2009} considers a setup where a source transmits updates that are distributed over a graph by a gossip network. An early attempt to study \gls{aoi} in gossip networks is \cite{JSelen2013}. Yates in \cite{yates2021gossip} provides \gls{aoi} analysis tools for gossip algorithms on network graphs, and the metric of version \gls{aoi} is defined therein, extended in \cite{yatesSPAWC21, BuyukatesJSAC2022}.

\gls{aos} is defined in \cite{ZhongISIT2018} to capture the freshness of a local cache. Moreover, the problem of how a local server allocates the refresh rate for each source to maintain overall data freshness given a constraint on the total refresh rate is also studied therein.

The challenges for \gls{aoi} optimization are also particularly relevant in the integration of vehicular and non-terrestrial networks in 6G: in the former, large quantities of data need to be exchanged while maintaining a very low \gls{aoi}, in a highly dynamic environment, and distributed \gls{rl} solutions, along with the use of learning-based models at the edge, can control the trade-off between the mobile network efficiency and the \gls{aoi} of the sensor data~\cite{sliwa2021client}. In the latter, the choice of the access scheme for traffic offloaded to satellites or drones is complicated by the higher propagation delay and the fast-changing network topology~\cite{zhang2021aoi}. Non-orthogonal schemes can be beneficial in some cases~\cite{gao2022non}, although the benefits are strongly dependent on the sampling process~\cite{chiariotti2022ran}. In these cases, repetition can also provide some additional reliability and improve \gls{aoi}: \gls{irsa}~\cite{munari2021modern} is a non-orthogonal solution that can allow large numbers of nodes to maintain a low \gls{aoi} without explicit coordination, using packet replication to protect the transmission from collisions. More complex retransmission schemes can also be used if packets are acknowledged, in which case more aggressive retransmission policies can be beneficial for \gls{aoi}~\cite{zhou2021performance}.

However, there is a tacit assumption in both \gls{aoi} and \gls{paoi} that the information should be fresh at any time. We may instead consider the case in which an application at the receiver accesses the information only at specific points in time, as introduced by the \gls{qaoi} framework~\cite{holm2021freshness}: this metric is similar to \gls{paoi}, but instead of sampling the \gls{aoi} only prior to a new packet reception, it does so when the application requests the information (i.e., when the information has the highest value). This transforms the setting from a push-based system to a pull-based one, where the application is dictating the transmission process. If the application works over discrete time intervals, then this leads to a better characterization of the information freshness as perceived by the application, and using it leads to very different choices in terms of system optimization~\cite{li2020waiting}. In the schematic from Fig.~\ref{fig:layermodel}, this would be equivalent to integrating information about the timing step of the estimator, monitor, or controller, as well as the sampling process at the source.

The last example of composite measures considers a sense-compute-actuate cycle where the system has a requirement regarding the maximum time between an event and the corresponding action. In this kind of wireless network controlled systems, Node 1 plays a dual role of sensor and actuator, and Node 2 is the remote controller. Node 1 sends the state of the system to Node 2, and Node 2 replies with a control command to Node 1, which acts accordingly. The \gls{aol} can then be defined as the \gls{aoi} over the two-way connection, which includes both the transmission of the system state from Node 1 to Node 2 and the transmission of the command from Node 2 to Node 1. In most systems, the command is a very short packet with a deterministic latency, so it simply adds a constant value to the one-way \gls{aoi}. If the command has a stochastic transmission latency or a significant size, the overall \gls{aol} is equivalent to the \gls{aoi} in a tandem system which involves the two directions of the connection, significantly affecting considerations for system optimization. A clear example of this difference is given by \gls{ar} systems, in which the visor transmits the image from a camera to a server. The server then renders virtual objects in the physical space and sends the content back to the visor: both the uplink and downlink flows can have a significant throughput~\cite{jia2018delay}, and the perceived delay between the user's movements and their effect on the virtual content depends on the \gls{aol}. In this case, Node 1 is at the same time the past anchor and the future timing reference~\cite{deSantAna2021AoL}. 

The definitions of \gls{aoi} and \gls{paoi} can readily be extended to the multi-source case, as well as to the case where the sources are scheduled by the destination node. For a more complete overview of the literature on \gls{aoi}, we refer the reader to~\cite{yates2020agesurvey}.

\subsection{Beyond Age of Information: Composite Measures and Data Quality}\label{ssec:beyond_aoi}
Coming back to our initial question of what is the right piece of information that should be transmitted at each time instant, a more complex set of timing measures arises when we aim at capturing both timing and other data quality attributes that define its significance for the system ultimate goal. Data quality has been broadly studied in the context of information systems with many different definitions and lists of attributes~\cite{cichy2019overview}. In any case, several of the desirable attributes for data quality are related to timing: freshness, currency, age, obsolescence, or staleness are often considered. Besides, data must be relevant, reliable, accurate, and complete. 

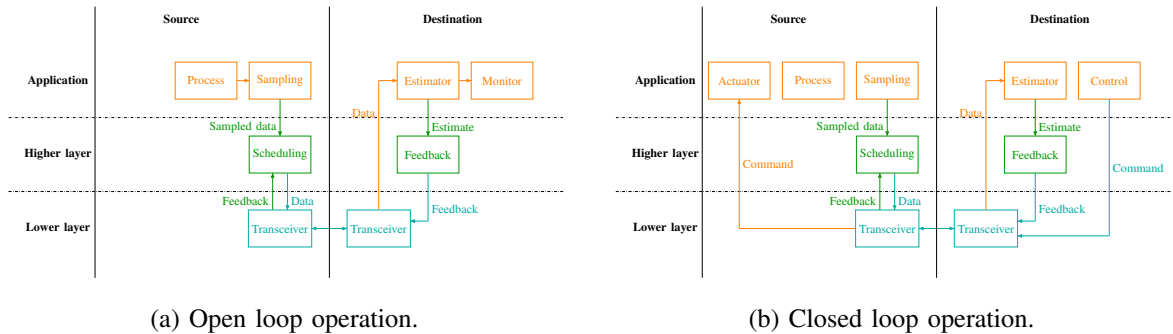
\begin{figure}[t]
 \centering
 \begin{subfigure}[b]{0.48\linewidth}
	    \centering
        \resizebox{0.95\linewidth}{!}{
        \input{figures/layer_model}}
        \caption{Open loop operation.}
        \label{fig:open_loop}
 \end{subfigure}
  \begin{subfigure}[b]{0.48\linewidth}
	    \centering
        \resizebox{0.95\linewidth}{!}{
        \input{figures/layer_model_closedloop}}
        \caption{Closed loop operation.}
        \label{fig:closed_loop}
 \end{subfigure}
    \caption{Transmission model in an \gls{aoi} (or beyond \gls{aoi}) optimization scheme.}
 \label{fig:layermodel}
\end{figure}

Integrating the data quality in our network design can be done by moving from measures based on statistical decisions, i.e., on the average \gls{aoi}, potentially measured at a specific point in time, or on the probability that its value is below a certain threshold, to measures based on statistical estimation, where we need to look more closely at the process that is being measured. Using this approach, several new composite metrics have been defined linking freshness and significance. The final objective is to design goal-oriented or \emph{semantic} communications~\cite{popovski2020semantic,kountouris2020semantics, uysal2021semantics} for networked intelligent systems able to optimize the use of resources. Fig.~\ref{fig:layermodel} shows a system model that can exploit this approach: while \gls{aoi} is an end-to-end metric at the higher communication layers, and its optimization requires no knowledge of the blocks at the application level beyond the sampling statistics at the source, we can extend the framework to the statistical estimation by introducing knowledge about the process statistics at the source and the estimation and tracking process at the receiver. A classic example is the Kalman filter: if the receiver uses this filter model, as in~\cite{mason2020adaptive}, it is possible to maximize the accuracy by taking the uncertainty on the estimate, which is provided by the filter itself, into account. We can look at Fig.~\ref{fig:open_loop} for a fuller picture of how such an open loop remote estimation system could work: unlike in simple \gls{aoi} minimization, the feedback module should be provided with statistics on the current estimates, so that it can make a better decision. Depending on the topology of the network and the capabilities of the nodes, we might consider the scheduling of updates to be entirely directed by the receiver. The receiver sends requests for new updates when it needs them, entirely directed by the source, which maintains an estimate of what the receiver knows and checks it against the actual state of the process, or distributed, with both nodes having partial knowledge and partial responsibility for the final decision on when and what to transmit.

The first example of these estimation-based metrics in the literature is the \gls{aoii}~\cite{maatouk2020AoII}, which extends the notion of fresh updates to that of fresh ``informative'' updates in statistical estimation, such that the age increases when the quality of the estimation of Node 2 about the process at Node 1 deteriorates. \gls{voi}~\cite{ayan2019age} is a metric that looks not only at the freshness of the new information being transmitted, but at its content, as the relative timing reference is based on statistical estimation, and the objective is to minimize the difference between the actual measured process and the one estimated by the receiver through the updates.

We can now consider the system in Fig.~\ref{fig:closed_loop}: in this case, the feedback loop is closed not only at the higher transmission layer, but also at the application layer, as the receiver exerts some control over the remote process by controlling an actuator. Teleoperation and remote control are standard examples of this kind of system, and in these cases, the main metric of interest is not the accuracy of the estimation, but its effect on the control performance. \gls{uoi} is a recent extension of \gls{voi} that considers this~\cite{zheng2020urgency}, making the objective not just the accuracy in measuring the process, but the stability of its control. If the controller needs to rely on the information sent over the wireless link, information that changes control decisions should be prioritized. 

We would like to emphasize that all these \gls{aoi} variants and other composite measures are metrics that can unify the past and future timing references. A generated packet starts its aging process immediately after its generation, so this is the past timing reference. In addition, systems can operate with \gls{aoi} or \gls{voi} thresholds, so that a new status update will be generated when the metric has reached a given value, which is a future time reference. When applied in different contexts, this metric can then provide a more holistic view of timing in communication systems.

\subsection{Protocols and Standardization Efforts} \label{sec:3gpp}
Besides academic research, there is a vivid interest and on-going work in the industry related to timing-aware designs for future communication networks. 

One example is the 5G technology: over the last decade, the \gls{3gpp} made a great effort to understand the most relevant use cases and applications from the so-called \emph{vertical domains}. This effort led to the initial classification into three generic services: \gls{embb}, \gls{urllc}, and \gls{mmtc}~\cite{popovski20185g}, whose scenarios and requirements are set out in~\cite{3GPPTR38.913}. However, these categories are insufficient to capture the complexity and intricacies of next generation of systems, including timing relations that go beyond the classical end-to-end latency and the reliability-latency couple. Therefore, \gls{3gpp} has continued the work to identify service requirements for new applications such as the factories of the future, cyber-physical control applications, utility grid protection, medical monitoring, and autonomous driving~\cite{3GPPTS22.261,3GPPTS22.104}. 
Three interesting timing metrics have been defined. The first one is the \emph{survival time}, which is the time that an application consuming a communication service may continue without an anticipated message. We notice that this is dependent on the application and the allowed set of sequence of failures. Referring to our example from Fig.~\ref{fig:TimingRefsIllustration}, one can understand the deadline $T_d$ as a survival time. 
A related concept is the \emph{watchdog timer}, used in control applications to automatically reset a device that hangs because of a software or hardware fault (or due to a delayed or lost packet when there is a communication network)~\cite{miremadi1995evaluating}. 
The second metric is the \emph{transfer interval}, which is in principle more relevant for periodic communication, but also applicable to scheduled aperiodic traffic. It is defined as the time elapsed between any two consecutive messages delivered by the automation application to the ingress of the communication system. This measure is related to the relative timing reference.
The third metric is the \emph{\acrfull{ttff}}~\cite{kruczynski1995time}, applicable to high-accuracy positioning and giving the time elapsed between the event triggering for the first time the determination of position-related data and the availability of position-related data at the positioning system interface~\cite{paonni2010performance}. This metric is related to a statistical decision with a past timing anchor.

Another interesting addition is the communication service reliability, which enlarges the \gls{urllc} reliability definition and it refers to the ability to provide the communication service for a given time interval but under given conditions. These conditions would include aspects that affect reliability, such as mode of operation, stress levels, and environmental conditions. Reliability may be quantified using appropriate measures such as mean time between failures, or the probability of no failure within a specified period of time.

Requirements and definitions for the system synchronization are also observed, with multiple time domains: the global time domain, used to align operations and events chronologically; and the working clock domains, i.e., for a machine of set of machines that physically collaborate. Different working clock domains may have different timescales and different synchronisation accuracy and precision. 
Analogously to a latency budget, there is a synchronicity budget with the time error contribution between ingress and egress of the 5G system on the path of clock synchronization messages. Current solutions to achieving fast and continuous synchronization~\cite{levesque2016sync} will be certainly not sufficient to satisfy the demanding timing relations of the future use cases. For instance, industrial automation scenarios typically involve multiple timing domains. 
Despite the challenge of integrating 5G into a \gls{tsn} synchronization network, the \gls{3gpp} has been working to make it feasible~\cite{godor2020}\cite{nasrallah2019}. Mechanisms for clock distribution are already included in Release 16, and the architectural solution has been consolidated in Release 17~\cite{3GPPTS23.501}. 
\gls{5gacia} was established in 2018 and aims at bringing all industrial and networking stakeholders together to accelerate the adoption of 5G technology in the industrial domains. One of the objectives was to ensure that the requirements are adequately addressed in 5G standardization and regulation, and this includes many timing-related dependencies. For example, we already mentioned the report in~\cite{5gaciakey}, where the time-scale categories are defined, whereas~\cite{5gaciatsn} describes the requirements and functional capabilities needed to integrate 5G with \gls{tsn}.  

Finally, timeliness is identified by \gls{3gpp} as an attribute for timing accuracy useful to quantify the end-to-end latency~\cite{3GPPTS22.104}. A message is considered in time if it is received within the timeliness interval given by the target value and the lower and upper bounds given by the allowed earliness and lateness. Besides, there can be a deviation or discrepancy between the actual time value and the target. This approach is useful to, e.g., applications where one of the nodes does not keep its own time, but interprets the message arrival as a clock signal. Nevertheless, no specific metrics are defined.  

Naturally, \gls{3gpp} is not the only entity working on timing requirements and standards, as several higher-layer protocols whose specifications are published by the \gls{ietf} also consider timing. The well-known \gls{ntp}~\cite{rfc5905} is an older example that takes network timing requirements into account to achieve clock synchronization between different computers. In this case, the two-way latency is the critical parameter: as the two endpoints need to establish a common clock, they cannot rely on knowing the latency, and the common assumption is that the path is symmetrical, i.e., the one-way latency is half of the overall \gls{rtt}.

Several other well-known protocols take time explicitly into account, using timing signals to trigger state changes and actions: a well-known example is the retransmission timeout in the \gls{tcp}, which is triggered if a two-way deadline is not met, i.e., if the acknowledgment for a packet is delayed by more than a certain time~\cite{rfc6298}. The \gls{rtp}~\cite{rfc3550} is another end-to-end protocol that takes timing into account, as it is designed for streaming media. \gls{rtp} packets are timestamped to compensate for jitter, playing each video frame or audio sample in the correct order and at the correct time, and synchronizing events in a manner similar to the example from Fig.~\ref{fig:LightingTrain}.

\section{Timing in Point-to-Point Communication Models}\label{sec:p2p}

We can now look at timing in point-to-point scenarios, i.e., when we have two endpoints communicating with each other. This is the simplest case of timing in networking, as information is exchanged one to one.

\subsection{Timing in Shannon's Communication Model}

\begin{figure}[t]
 \centering
 \input{./figures/shannon.tex}\vspace{0.3cm}
 \caption{Shannon's communication model annotated by layering.}
 \label{fig:ShannonLayered}
\end{figure}
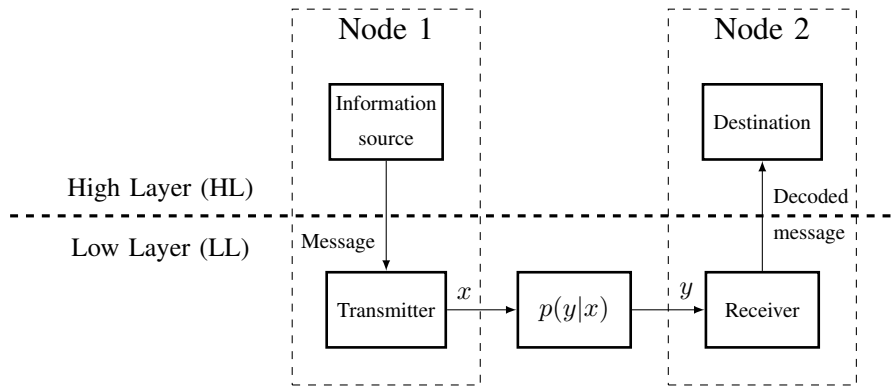

We start from the well-known Shannon communication model~\cite{Shannon48}. A variation on this model is depicted on Fig.~\ref{fig:ShannonLayered},
with two modifications from the original model in~\cite{Shannon48}: first, the noise source is absorbed within the conditional probability distribution $p(y|x)$ between the input $x$ and the output $y$ of the noisy communication channel, and second, the modules of the model are placed within the context of a layered model of a communication protocol, consisting of only two layers: \gls{hl} and \gls{ll}. It will be seen that these two layers are sufficient to introduce different notions of timing and their interaction through the layers.  

The elementary unit in Shannon's model is the \emph{channel use}, specified through the probabilistic relation $p(y|x)$ between input $x$ and output $y$. The sequence of channel uses $x_1, x_2, \ldots x_n$ are causally connected, such that, for $i<j$, $x_i$ takes place before $x_j$. Nevertheless, strictly speaking, timing is \emph{absent} from Shannon's communication model. To elaborate, it should be noted that the model does not specify \emph{(i)} that the time interval between two adjacent channel uses should be constant, and \emph{(ii)} what is the value of a time interval between any two adjacent channel uses.

Shannon's model is a mathematical model and in order to connect it to a physical quantity, such as time, we need to specify additional parameters.
We emphasize these points, as they go against the common view on communication systems, which assumes \emph{a priori} that there is a constant time interval $T$ between two adjacent channel uses, such that a data packet that requires $n$ channel uses takes a time $nT$. Indeed, the norm in communication systems is that channel uses occur periodically, as a result of a certain sampling process. If periodic Nyquist sampling is assumed, the period between two samples is $\frac{1}{2B}$, where $B$ is the used bandwidth. The time $T_n$ that corresponds to $n$ channel uses is then
\begin{equation}
    T_n=\frac{n}{2B}.
\end{equation}
Nevertheless, once a time duration is associated with $T_n$, for each $n>0$, then latency of a given packet transmission is directly defined\footnote{Strictly speaking, the duration of a packet with $n$ channel uses is $\frac{n-1}{2B}$, counting that the first channel use occurs at time $0$, but considering that $n\gg 1$ we will neglect this fact throughout the discussion.} as $T_n$. This is one common way in which timing enters the Shannon communication model.

The problem of transmission over a noisy channel, referred to as \emph{channel coding problem}, deals with the design of the \gls{ll}. In doing that, this problem is solved under a simple abstraction for the operation of the \gls{hl}: the information source selects uniformly randomly one of the $M$ possible messages and this message should be recovered at the destination, which is also a \gls{hl} module. This simple abstraction is a consequence of the separation between the channel and source coding, where the latter is assumed to deliver perfectly compressed messages to the transmitter. Furthermore, it is assumed \emph{a priori} that there is data passed from the source to the transmitter, such that the transmitter is always actively transmitting and having an idle channel is not part of the model\footnote{In fact, an idle channel can be seen as a specific type of transmitted symbol. Therefore, accounting for this requires to change the communication model.}.

Next, we discuss how Shannon's model can be used to represent the different options from the statistical framework from Sec.~\ref{sec:framework}. Note that, in this communication model, the message sent by the transmitter is already selected, encoded, and ready to be sent when the first channel use occurs. This involves  at least the following tacit assumptions: 
 \begin{enumerate}
\item The receiver decodes the message instantly;
\item There is no additional latency due to protocol interactions;
\item Both the sender and the receiver are certain that the data selected by the sender is useful and timely for the receiver.
\item There is always data available.
\end{enumerate}

We can conclude from this discussion that the only parameters that influence timing in Shannon's model under the assumption of Nyquist sampling are the number of channel uses and the bandwidth. Note that, for given $p(y|x)$, the channel capacity has an operational meaning by transmitting over asymptotically many channel uses. This may be interpreted as a communication that takes asymptotically long time or, alternatively, asymptotically large bandwidth. However, while increasing the bandwidth indeed leads to lower latency, the assumption of perfect compression means that the number of channel uses is already minimized, and thus is generally not a free parameter unless we allow for distortion, as discussed later for the statistical estimation case. As such, the Shannon model has a deterministic latency, and can be used for future timing references (e.g., it is easy to compute whether it will meet a deadline). However, this is not the case for timing metrics using a past anchor related to a physical process, such as \gls{aoi}, since the process of sampling of the data has been abstracted from the model. Finally, the modeling of a relative time reference
is trivial for the case when Node 1 determines the reference, but not possible in general, as the model has already predefined the role of a sender and recipient (e.g., in Fig.~\ref{fig:ShannonLayered}, Node 2 cannot be a transmitter).

In order to consider the statistical characterization of the timing measures, we first examine the statistical decision process. The latency-reliability function of a given channel can be determined by using finite-blocklength theory \cite{Polyanskiy10}. Specifically, once bandwidth and $T_n$ are fixed,
then the number $L$ of channel uses within the time interval $\Delta t$ is also fixed. For a given size $D$ of a data packet, the probability in~(\ref{eq:LatencyReliability}) can be expressed in terms or the probability of success for a given data rate, expressed as bits per channel use, and this is precisely the type of result that can be obtained using finite-blocklength theory. 

We can now examine the statistical estimation case. Here, we need to consider the source encoding aspect of the model from Fig.~\ref{fig:ShannonLayered}, in order to show the quality of the estimated value at the receiver at the end of a given interval $\Delta t$. More precisely, the characterization of the quality of estimation that Node 2 can get after a certain number of channel uses is a subject to joint source-channel coding.

Some of the assumptions for the Shannon model, listed above, can easily be  generalized. For example, instead of assuming that the receiver decodes the message instantly, we can assume that there is a certain decoding time $T_d$, potentially dependent on the actual received signal (e.g., resulting from a number of iterations of a belief propagation decoder due to noisy reception) as well as on the content of the data (e.g., with the use of unequal error protection). 

Assumptions 2 and 3 follow from the simplified functioning of the \gls{hl} in Shannon's model. The transmitter at the \gls{ll} obeys the command to transmit the data, not questioning its effectiveness or ultimate interpretation by the receiver. In principle, the model does offer the freedom to perform source coding and thus address the problem of statistical estimation. However, to in order to characterize the source coding operation in terms of timing, one needs to enrich the model by introducing a time processing/computing related to the source coding operation.

\subsection{Timing in a Two-Way Communication Protocol} \label{sec:two-way}

\begin{figure}[t]
 \centering
 \input{./figures/twoway.tex}\vspace{0.3cm}
 \caption{Two-way communication model with two layers. HTR$i$ is the HL transceiver at node $i$, LTR$j$ is the LL transceiver at node $j$. }
 \label{fig:TwoWayHLLL}
\end{figure}
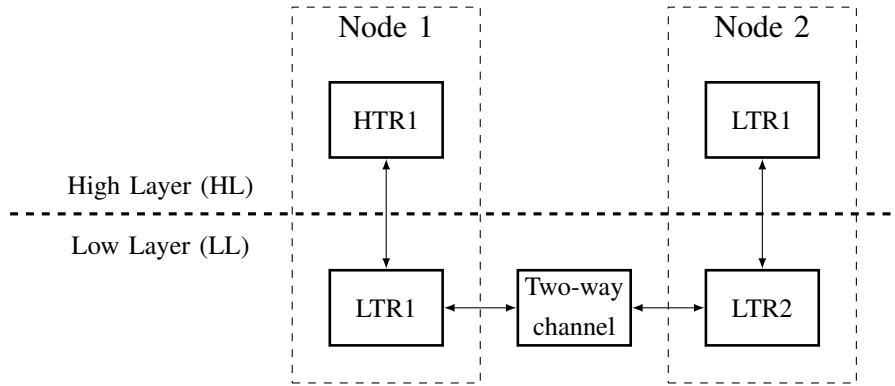

In order to introduce more complicated protocol interactions, the one-way communication model from Fig.~\ref{fig:ShannonLayered} is generalized with the two-way model from Fig.~\ref{fig:TwoWayHLLL}. In this model, each module is a \emph{transceiver}, as it can both transmit and receive. The model loses the simplicity and the mathematical elegance of Shannon's one-way model, but it is suitable to introduce protocol interactions, without the ambition to have a consistent model that can mathematically describe the ultimate efficiency bounds, as it is the case with the notion of channel capacity. We assume a \gls{tdd} system in which the time division is followed ideally by both nodes, i.e. when Node $i \in \{1,2\}$ is in transmit state, then Node $j \neq i$ is always in a receive state. Note that in this model we are not talking about individual channel uses, but rather about a duration of a transmission; in that sense, the duration can also accommodate the idle time that occurs due to the switching in a \gls{tdd} system. 
 This model can be generalized to other duplexing scenarios, but is not essential for the discussion on timing.   

In this model, Node $1$ is the originator of the useful data (shortened to ``data'' when there is no danger of confusion). More specifically, each message is assumed to originate at the \gls{hl} transceiver HTR1 and is passed on to the \gls{ll} transceiver LTR1 to be sent over the noisy channel. To make timing part of the channel definition, let us consider some simple protocols that can take place in this model and start with a protocol for acknowledged transmission. Assume that Node 1 is a sensor that can measure some time-variant quantity in a \gls{cps}. HTR1 samples the system at time $t=0$, processes the data and at passes the data of size $D_1$ on to LTR1\footnote{We can add a processing delay in HTR1, but we keep it simple for the moment.}. This data is immediately sent to LTR2. Upon decoding the data, LTR2 passes it on to HTR2 and simultaneously sends and acknowledgement of size $D_a$ to LTR1. Upon receiving the acknowledgement successfully, LTR1 immediately informs HTR1. Provided that the transmission of both data and acknowledgment are successful, the acknowledgment is received after one \gls{rtt}, $t_{\text{RTT}}$, which can be computed using the same tools as in the one-way channel. This leads us to the first characterization of latency in a two-way communication protocol as a statistical decision process, where the latency is defined as the time from data transmission until reception of the acknowledgment. However, while it was sufficient to fix the bandwidth and the symbol period $T_n$ in the one-way channel, the two-way channel also requires us to specify how the allocated symbols are split between data transmission and for acknowledgment. Let us denote the fraction of symbols used for the data transmission by $k$, so that when there are a total of $L$ channel uses within the time interval $\Delta t$, the data is transmitted with rate $R_1=\frac{D_1}{kL}$, and the acknowledgment has rate $R_a=\frac{D_a}{(1-k)L}$. The latency-reliability function can then be obtained as the convolution of the one-way latency-reliability functions of the data and acknowledgment transmissions computed for $R_1$ and $R_a$, respectively. 

Similar to the Shannon model, the protocol for acknowledged transmission is a \emph{push-based protocol}, where the originator of the data (Node 1) knows exactly which data and when it is requested by the destination (Node 2). The model can be also used for \emph{pull-based protocols}, where Node 2 initially sends a request of size $D_r$ to Node 1, and in turn, it receives the data it requested. In this case, the timing quantities of interest are measured with respect to the moment when the request is sent by Node 2, not with respect when the data is sent by Node 1. The model can also be extended to include retransmissions at LTR1, where the data is retransmitted until an acknowledgement is received. In this case, $\Delta t$ represents a random multiple of $t_{\text{RTT}}$ representing the number of transmissions required to get a successful data followed by a successful acknowledgment transmission, taking into account potential use of retransmissions and \gls{harq}.

To characterize the statistical estimation, consider the case where Node 2 estimates the state at Node 1. Compared to the one-way model, the main difference is that the feedback can be used by Node 1 to decide what to transmit next, or how to encode the data in the next transmission. This may for instance be the case in training of a machine learning model, where Node 2 keeps requesting data until the test accuracy has converged or is sufficiently low. These benefits are also pronounced in pull-based systems, where Node 2 can use side information, such as confidence intervals, to decide which information to request in order to minimize the estimation error.

So far, we assumed that Node 1 had the data that was of interest to Node 2. In some cases, a more reasonable model is to assume that Node 1 and Node 2 are interested in exchanging data and coordinate decisions. This could for instance be the case in a factory automation scenario where two robots need to coordinate in order to solve a task. We defer the discussion of this case to Sec.~\ref{sec:dist_concensus} where we consider the more general problem of distributed consensus.

A practical example of this kind of system is given by \gls{ntp} synchronization: in this case, the data that Node 1 transmits is itself a timing reference. We note that, in a general case that does not involve a Shannon channel with deterministic packet transmission times, such as most wireless systems, the jitter in the network prevents the two nodes from knowing the one-way latency, unless they already have a common timing reference. \gls{ntp} assumes that the latency is symmetrical, dividing the \gls{rtt} in two. As such, its precision is limited by the network latency. Naturally, the same considerations hold for timestamped data if the two nodes do not have a precise common reference: by determining the \gls{rtt}, the two nodes can agree on a shared timeline of events, but the order of events that occur at the two nodes within one \gls{rtt} of each other is impossible to determine in the absence of any information on the one-way latency.

\subsection{Timing and Freshness of Updates}
The \gls{aoi}, which is one of the simplest timing metrics, extends latency in that it accounts for the characteristics of the source (information generator) as an entity regarding freshness.

The baseline one way communication model consists of Node 1, which observes a certain process that evolves\footnote{A generalization is to consider that the process changes over space and time, such as the temperature in a smart factory depending on the operations of machines that can create heat.} over time and of Node 2, which is interested in having the latest status of that process. For that, Node 1 needs to measure (sample) the process, generate an update at HTR1 that describes the status of the process and forward it to LTR1 to be sent to the LTR2 and, ultimately, HTR2. The requirement for having the latest possible update at Node 2 has an impact on both the data generation process at Node 1 (i.e. when sampling/measuring process occurs) as well as the data transmission process at LTR1. Assume that HTR1 forwards an update $D_{1}$ to LTR1 at time $t_1$. At a time $t_2>t_1$, HTR1 generates a new update $D_2$, but when it communicates (internally within Node 1) with LTR1, it finds out that the data $D_1$ has not yet been transmitted. Then, one action is to request LTR1 to purge $D_1$ and replace its transmission with $D_2$, such that communication resources are utilized for the latest update. However, this heavily depends on the purpose of communicating the status updates. In case one is interested in tracking as well the past of the process, we may still need to transmit the freshest status update. Instead of purging the older ones though, LTR1 stores them in a queue and transmits them when no more urgent transmission is required. Since these updates are timestamped, they can be reordered upon reception at LTR2 to provide the required knowledge about the past. Furthermore, if there is a non-negligible communication delay among LTR2 and LTR2, we may consider that LTR2 will forward the freshest received update to HTR2 and then the rest will follow. This can be crucial in cases when there is a need for actuation at Node 2. 

The concept of \gls{aoi} and information freshness can be extended to two-way communication: Node 1 cannot know how fresh is the information at Node 2 unless Node 2 provides feedback, which means that there is a two-way communication in place, and \gls{aol} accounts for the feedback loop as well. Another interesting two-way scenario is the one in which Node 1 is a sensor/actuator, while Node 2 is an edge controller. Node 1 sends a state to Node 2 and expects back a command that needs to be actuated. In this case, the correct system operation and stability depends on the \gls{aol}, defined as the age of the two-way loop and measured from the moment the state is sent until the command is received~\cite{deSantAna2021AoL}. In general, one can define various composite measures of freshness taking into account the states and the data reception at both nodes, as we described in Sec.~\ref{ssec:beyond_aoi}.

\subsection{Timing in a Cascade of Modules}
\begin{figure}[t]
 \centering
 \begin{subfigure}[b]{0.9\linewidth}
	    \centering
        \input{./figures/cascade_a}
        \caption{The processing time of each module depends only on the data.}
        \label{fig:cascade_indep}
 \end{subfigure}
  \begin{subfigure}[b]{0.9\linewidth}
	    \centering
        \input{./figures/cascade_b}
        \caption{The processing times of computation and compression are coupled through control metadata.}
        \label{fig:cascade_meta}
 \end{subfigure}
  \begin{subfigure}[b]{0.9\linewidth}
	    \centering
        \input{./figures/cascade_c}
        \caption{Computation and compression are combined in a single module and their processing times are inseparable.}
        \label{fig:cascade_dep}
 \end{subfigure}\vspace{1cm}
 \caption{An example of information processing of through a cascade of modules.}
 \label{fig:CascadedModules}
\end{figure}
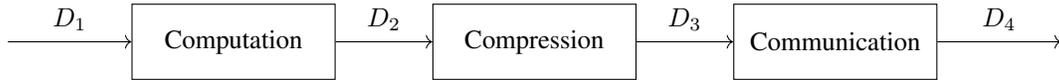
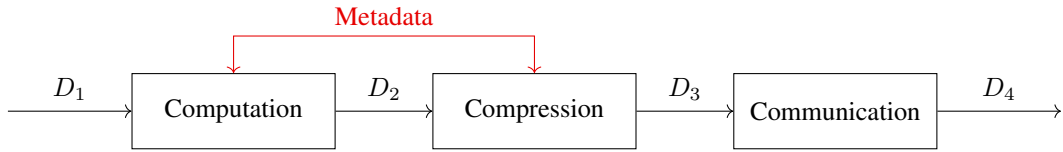
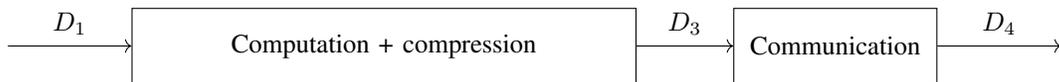

In the definition of information freshness, we have involved a cascade of modules from two layers, the higher layer doing the sampling and using the sample and the lower layer responsible for transport of the sampled data. In general, end-do-end information processing is a sequential process that involves different operations, such as computation, compression, encoding, baseband processing, etc. This can be conveniently represented through a cascade of modules, exemplified in Fig.~\ref{fig:CascadedModules}. It is easy to note how this model corresponds to the general system depicted in Fig.~\ref{fig:layermodel}, and can be applied in the scenarios we discussed with respect to the new timing metrics in the literature. It should also be noted that the communication module itself can have multiple sub-components, such as a wireless access link and a core network connection.

Let us take at first the case on Fig.~\ref{fig:cascade_indep}, in which the processing time of a given module depends only on the input received from the previous module. Recalling the latency-reliability function $\Pr(\Delta t \leq \tau \mid  D)$ from~\eqref{eq:LatencyReliability}, we can now interpret $D=D_1$ as the input data to the first processing module (computation). The total timing budget is given by the interval:
\begin{equation} \label{eq:SumTimeCascade}
    \Delta t = \sum_{i=1}^M \Delta t_i
\end{equation}
where $\Delta t_i$ is the contribution of the $i$-th module. For the model on  Fig.~\ref{fig:cascade_indep}, the processing time $\Delta t_i$ depends only on the input data $D_i$; that is, given $D_i$, then $\Delta t_i$ is conditionally independent of the other $\Delta t_j$ where $j \neq i$. This model is the basis for optimizing the latency budget: For example, one can increase the processing time of the compression module in order to get a better compression, which would decrease the size of $D_2$ and lead to a potentially shorter $\Delta t_3$.

Fig.~\ref{fig:cascade_meta} depicts an two-way exchange of metadata/control information that coordinates the processing of modules for computation and compression\footnote{In this example we are illustrating a possible coupling between computation and compression. In general, compression can be coupled with communication or all three operations can be joined into a single module.}. The cumulative processing time is again given by~\eqref{eq:SumTimeCascade}, but now $\Delta t_j$ for $j=1, 2$ is not only dependent on $D_j$: the joint distribution of $\Delta t_1$ and $\Delta t_2$ also depends on the exchanged metadata. For example, the computation module may signal to the compression module that current data has a higher priority, which will change the processing in the compression module and lead to a lower $\Delta t_2$. In principle, the timing performance attained for coupled modules on Fig.~\ref{fig:cascade_meta} should never be worse than the one for Fig.~\ref{fig:cascade_indep}. Nevertheless, when working with very short time scales, one needs to take into account the timing performance of the metadata/control exchange as well. 

Finally, in Fig.~\ref{fig:cascade_dep} the total timing budget is given by:
\begin{equation} \label{eq:SumTimeCascade2}
    \Delta t = \Delta t_{12}+\Delta t_{3}
\end{equation}
where $\Delta t_{12}$ is the time consumed by the joint computation/compression.  For a well-defined scenario, Fig.~\ref{fig:cascade_dep} can be optimized to achieve the best timing performance, i.e. it can be ensured that $\Delta t_{12} < \Delta t_{1}+\Delta t_{2}$. This is because any operation regime that can be attained in Fig.~\ref{fig:cascade_meta} can be attained in Fig.~\ref{fig:cascade_dep}, but not vice versa. 

The contrast between Fig.~\ref{fig:cascade_meta} and Fig.~\ref{fig:cascade_dep} reflects the ever-present trade-off between architecture and performance. Fig.~\ref{fig:cascade_meta} reflects an architecture that can scale and proliferate, such as the Internet, where the interaction among black boxes takes place through well-defined interfaces. This is also the approach of \gls{oran}~\cite{gavrilovska2020cloud}, which allows the wireless system to be built based on components with open interfaces. What is uncertain in terms of performance is whether the specification of the interfaces between the black boxes, along with the timing performance of their interaction, is capable to offer superior timing performance. However, given the interfaces, there is a broad base of competitors that can offer new processing algorithms and smart interactions through those interfaces. One important feature to achieve this goal is synchronization: the system should operate under a global time domain and distribute the working clock accordingly, as explained in Section~\ref{sec:3gpp}. At the opposite side is a solution fully implemented by a single vendor, which can optimize the interactions and the timings of different operations beyond the limitations of the open interfaces. However, it is uncertain at the time of design whether the optimization is versatile enough to support all future timing requirements.

\section{Timing in Networking Models}\label{sec:net}
The features we discussed above were all related to a point-to-point model, which is conceptually simple, as the two actors' roles are clear. If we extend the framework to a \emph{network} of actors, we have three principal models:
\begin{enumerate}
    \item \emph{One-to-many} transmission: in this case, a single transmitter needs to relay information to multiple receivers. In more classical terms, broadcast or multicast applications follow this model;
    \item \emph{Many-to-one} transmission: in this case, an aggregator node receives updates from multiple sources. This can represent, for example, a remote estimation or control process, in which a central monitor gathers data related to a complex process from multiple sensors;
    \item \emph{Many-to-many} transmission: this is the most general case, in which there are multiple transmitters and receivers.  
    Examples within this category are updating of a DLT, a ``swarm'' scenario with coordinated robots or drones or distributed learning networks.
\end{enumerate}

In the following, we elaborate on representative examples of the many-to-one category (Section~\ref{sec:dist_concensus}) and the many-to-many paradigm (Sections~\ref{sec:dist_concensus}-\ref{sec:edgeinference}).

\subsection{Model for Networked Control beyond point-to-point} \label{sec:networked}

Networked control systems refer to systems where multiple devices exchange information with the aim to coordinate some action that requires precise control and frequent feedback. Control systems can be centralized, as is often the case in industrial manufacturing scenarios, or decentralized, e.g. a power distribution system comprising distributed energy sources and loads that need to be controller to maintain stability of the overall system~\cite{guerrero2012advanced}. Note that although sensors, actuators and controllers may be distributed, we limit the discussion in this section to the case where the control actions are given from a central entity, and dedicate the next section to the problem of distributed consensus that for instance arises in multi-agent systems.

In the centralized case, the systems usually fall into the category of real-time systems, and follow a star or ring topology with a single controller that receives samples from sensors and sends directions for actuators~\cite{chowdhury2017fast}. The timing requirements are dictated by the controller, which is typically executed periodically in sense-compute-actuate cycles, during which the controller receives feedback from sensors, computes new actions for actuators, and sends the actions to the actuators. This way, the communication link provides the means to close the feedback loop and to synchronize the components. The tight integration of communication and computation enforces a deterministic time schedule of the communication by reducing the need of queuing, dynamic scheduling, etc. 

To put this into the statistical framework, let us consider the timing from the point of view of the controller, which dictates the control process. When the controller is about to compute the control actions, the most critical timing references are the times at which the current sensor readings, which must be recent to allow the controller to pick an accurate control action. In particular, if the readings are dated, the controller needs to predict the sensor state based on what it already knows (i.e. operate in open-loop mode), which is likely to be associated with a high amount of uncertainty. This timing perspective can be described using the \gls{aol} and its variants sampled at the discrete instants where the controller is about to compute its next control actions, underlining the observation mentioned earlier that the information needs to be delivered on-time as opposed to in-time.

So far, we have assumed that the sensors and actuators are directly connected to the controller in a star or ring topology. However, the topology could also be a completely virtual overlay network that is implemented on top of a general infrastructure such as the Internet. This situation arises for instance in power distribution systems with separate power and communication infrastructure, teleoperation, and the Tactile Internet~\cite{simsek20165g}, where there is a physical distance between the sensors and actuators on one side and the controller and the operator on the other. The increased distance between the components, as well as the fact that the infrastructure is shared with other (unknown) users, makes precise control over the timing more challenging. One way to compensate for this is to reduce the frequency of the cycle, provided that the controlled process allows for it, or to adopt predictive strategies~\cite{farajiparvar2020brief} that attempt to estimate the state of the remote system and revise control commands accordingly. 

However, a more common strategy is to delegate part of the control to controllers close to the actuators, so that the system forms a hierarchical control system in which the primary (high-level) controller is responsible only for supervising the secondary controllers~\cite{shafiee2013distributed}. In this case, the timing perspective becomes more complicated, as it is not sufficient to consider timing from the perspective of the primary controller, but also of the secondary controllers, which should adapt to the state of the primary controller. Consequently, it is useful to adapt the same hierarchical structure when considering the timing anchors: the primary controller's reference is both the originator of its overall control strategy, e.g. an operator, and the state of the secondary controllers. Similarly, the references of the secondary controllers are both the state of the primary controller and the states of the sensors and actuators that it directly interacts with.

The hierarchical control structure can be generalized further to a fully decentralized control system where sensors, actuators and controllers are interconnected as a mesh network. From the perspective of a single device, the timing references are given by the states of all other devices. However, the overall objective carried out by the devices may have a deadline, as is the case in self-driving cars that are coordinating to prevent an accident. When the devices are acting independently towards reaching the overall objective the problem is characterized as a distributed consensus problem as discussed next.

\begin{itemize}
    \item \emph{Past Timing Reference}: In this case, the \gls{aoi} and \gls{aol} are the most relevant metrics. Having data from multiple sources, or even having multiple controllers coordinate, complicates the definition of a single metric, but common strategies are either to look at the average or to consider the last stragglers, adopting a more careful approach.
    \item \emph{Future Timing Reference}: If we know that the controllers have compensation systems for latency, as we discussed above, there is a limit to the maximum latency that they can compensate without significantly degrading the system's performance. We can then set a deadline from the moment the data are generated to the moment they arrive to the controller or controllers, or from the moment the data are generated to the execution of the subsequent control command. 
    \item \emph{Relative Timing Reference}: We can also consider the control system itself in the timing computation, setting objectives related to the control performance. In this case, the \gls{aoi} of measurements is substituted by the \gls{uoi} or \gls{aoii}, as we have explained in Sec.~\ref{ssec:beyond_aoi}.
\end{itemize}

\subsection{Distributed Consensus and DLT}\label{sec:dist_concensus}

\begin{figure}[t]
    \centering
    \input{figures/consensus.tex}\vspace{0.3cm}
    \caption{Simple example of a distributed consensus.}
    \label{fig:consensus}
\end{figure}
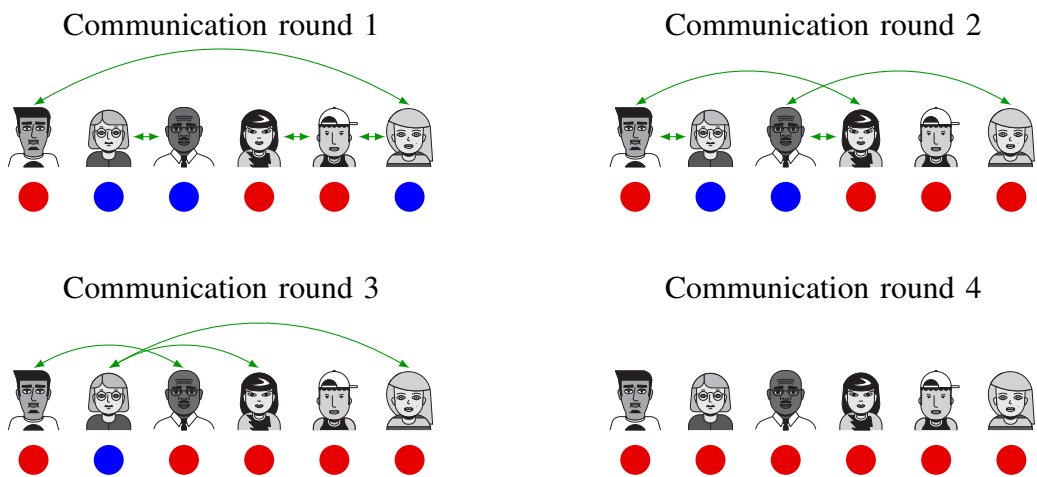

Consensus is a well-known problem in distributed systems, in which multiple nodes need to arrive at the same conclusion over a measurement or a future action by exchanging messages over a constrained communication system as a means to jointly achieve a global objective. The problem is exacerbated if a subset of the nodes are faulty or actively malicious, as the system needs to protect itself from incorrect or harmful information. A simple example of consensus is depicted in Fig.~\ref{fig:consensus}, in which 6 people (voters) must decide on casting a red or a blue vote. In each round, a maximum of two other voters can be contacted using unicast communication. Looking at the first round, voter 6 receives information that voters 1 and 5 will choose red, so being in minority she decides to change her vote to red. Instead, voters 2 and 3 exchange messages which reinforce their intention to vote for blue. 
In the figure, three rounds of communication allow the voters to agree on voting for red, but the solution and the time until it converges depends on the availability of communication resources and on the initial preferences of each person.

A canonical example of consensus in the presence of untrustworthy nodes is the Byzantine Generals problem~\cite{Lamport1982Byzantine}, where several divisions of the Byzantine army are camped outside an enemy city, each division commanded by its own general. The commanding general must decide on a plan of actions and communicate it to the other generals to be carried out in unison. However, there might be one or several traitors (including the commanding general itself) that disseminate false information or are otherwise unreliable. 

There are many practical examples of consensus in distributed computing~\cite{qin2016recent} and multi-agent systems. 
A prime example is for implementing distributed management of road intersections, for increased efficiency and safety. The idea is that the cars communicate to each other to coordinate their actions (speed, direction). The consensus is achieved through an iterative process where the location-aware vehicles select conflict-free trajectories that minimize their travel time. Besides the non-idealities of the communication channel, another limitation to take into account is the deviation of the measurements from the true location. %
A second example is represented by \gls{dlt}. Although the rise of blockchain and \gls{dlt} has been driven by cryptocurrency and supported by wired networks, there are more and more envisioned use cases in existing or future wireless networks~\cite{kuo2018}. Generally speaking, the DLT enables consensus among a group of geographically distributed nodes without a central controller. Examples in the wireless realm include ad-hoc networks and clusters of small cells. As all transactions need to be confirmed by a qualified majority of the nodes (weighted either through proof of work or other mechanisms such as proof of stake) before they are inserted in the ledger, timeliness is crucial. 
Furthermore, \gls{dlt} is designed to be resistant to malicious nodes, as long as they do not represent a majority, thus placing the problem in the ``Byzantine generals'' category.

Consensus in these scenarios is communication-heavy, and defining timing reference(s) in this context is not trivial. The simplest way to measure consensus is by taking a \emph{bird's eye} view of the network, i.e., considering consensus to be achieved at the instant when the last node required for the majority gets the necessary information. However, this view is often unrealistic in real networks, as it requires ideal communication links and full synchronization among nodes. Therefore, a full consensus can be defined when all nodes have been informed, taking the point of view of the last straggler node to confirm that it received and accepted the transaction: in game-theoretic terms, all agents must have complete information on the state of the system, i.e., not only must the consensus state be common knowledge, the fact that it is common knowledge must also be common knowledge. A relaxation to this definition would be to declare consensus once the node that started the update is informed that consensus is achieved.

Gossip networks, in which nodes propagate information generated by any of the others~\cite{yates2021gossip}, are a useful model for the scenarios we outlined above. A real application of gossip networks is the distributed ledger, in which a majority of nodes (often weighted by their computing power, or their stake in the transaction itself) must agree on a transaction before it can be inserted in the ledger. The meaning of consensus depends on the mechanism used for the ledger: classical blockchains use a mining process such as the well-known and resource-intensive proof-of-work~\cite{rovira2019optimizing}, while HyperLedger uses a lighter method that is based on simulating transactions against the local ledger~\cite{kim2020age}. In general, the hard consistency requirements before a transaction can be confirmed make distributed ledgers a very complex scenario in terms of timing, but clarify the timing anchor: the delay in a transaction is the time difference between the instant a transaction is initiated and the instant in which the originator node is updated with the block confirming that the transaction is registered in the distributed ledger.
\begin{itemize}
    \item \emph{Past Timing Reference}: In this case, we consider the consensus latency in a network, using the initial message as an anchor and computing the time until consensus is achieved. As we discussed above, the definition of consensus can be tricky, and different metrics can be devised depending on the precise objective of the latency.
    \item \emph{Future Timing Reference}: In the same way, the design of the system can be oriented at guaranteeing a maximum latency, using deadlines and allocating communication resources to the nodes so as to meet the deadline.
    \item \emph{Relative Timing Reference}: In a distributed ledger, the consensus latency includes both the communication latency and the duration of the consensus mechanism, which can be significant e.g. in proof-of-work systems, and the confirmation instant can have different definitions. This means that a model of the consensus mechanism must be included in the computation of latency or age, making the timing reference tied to this process.
\end{itemize}

\subsection{Timing in Distributed Machine Learning}
While the rise of cloud computing caused computing to become more centralized in 2010s, recent years have witnessed the opposite trend of pushing computing to the edge of a local network (i.e., devices and powerful computers near base stations), called \emph{\gls{mec}}. Several factors that drive the paradigm shift include the availability of powerful processors for both devices and servers, the emergence of latency critical applications, the issue of network data congestion, and the concerns over data privacy. Among many others, the training of \gls{ai} algorithms is an important application of \gls{mec}. In this subsection, we will discuss distributed machine learning (also called edge learning), while its use, called edge inference, is to be discussed in the next subsection. 

\subsubsection{Principle of Distributed Learning} 
Distributed machine learning refers to the distribution of a learning task over multiple edge devices to leverage either their data or computational resources or both. A learning algorithm usually attempts to minimize (or maximize) some function $L(\mathbf{w})$ of the model-parameter set $ \bm{w}$, referred to as loss function. Thus, a learning task involves finding the  optimal model $ \bm{w}^\star$ that solves an  optimization problem: $ \bm{w}^\star =\arg\min_{\bm{w}} L(\mathbf{w})$. 

\begin{figure}[t]
    \centering
    \input{figures/model_update.tex}\vspace{0.3cm}
    \caption{Federated learning system and its operations.}
    \label{fig:FL_Sys}
\end{figure}
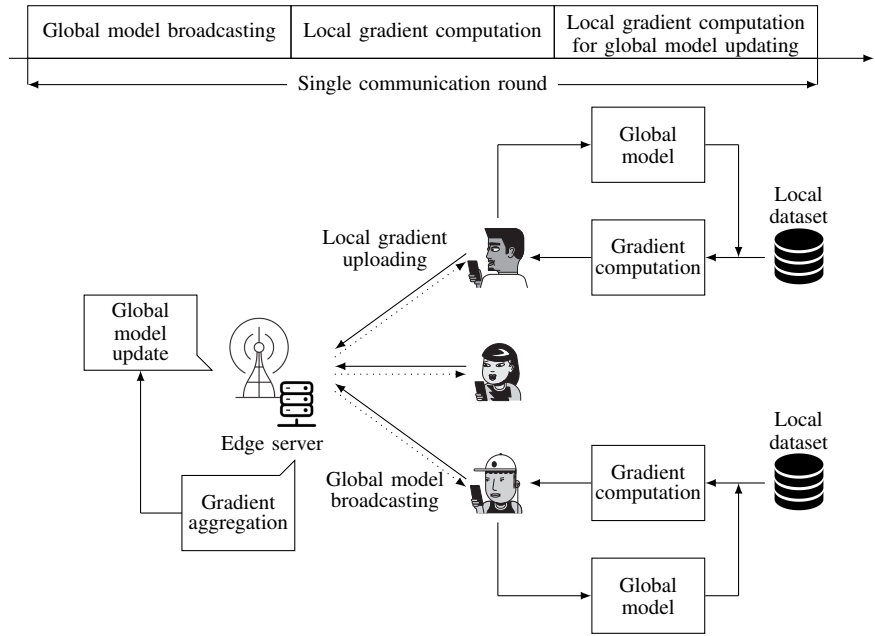

There exist various approaches to do this, designed for different purposes. Perhaps the most popular approach is \emph{\gls{fl}}, that solves the mentioned optimization problem by implementing the \gls{sgd} algorithm of in a distributed manner. Its key feature is the avoidance of direct data uploading, allowing the exploitation of users' data while preserving their privacy. The \gls{fl} system and operations are illustrated in Fig.~\ref{fig:FL_Sys}. The \gls{fl} iterative algorithms comprise multiple communication rounds. At the beginning of each round, say round $n$, a server broadcasts the global model to all devices for distributed model/gradient estimation, described as follows. The gradient corresponding to gradient descent on the function is called a \emph{ground-truth gradient}. Each device estimates the ground-truth gradient, $\nabla L(\bm{w}^{(n)})$, using its local dataset. The result is a \emph{local gradient} that is a noisy version of the ground truth. Upon the completion of local computation, each device uploads its local gradient to the server. Alternatively, each device updates the downloaded global model by performing multi-round gradient descent locally and  then uploads the resulting \emph{local model} to the server. To suppress the estimation noise, the server aggregates (i.e., averages) the local gradients (or local models) and applies the aggregation result to update the global model, completing the round. Let $g(n, \bm{w})$ be the global (aggregated) gradient in the $n$-th round, which is then applied to updating the global model based on gradient descent:
\begin{equation}
\bm{w}^{(n+1)} = \bm{w}^{(n)} - \mu g(\bm{w}^{(n)}). 
\label{Eq:global_update}
\end{equation}
In the case in which the distributed data are \gls{iid}, $g(n, \bm{w}^{(n)}) = \mathsf{E}[\nabla L(\bm{w}^{(n)})]$, and its variance from the ground truth is inversely proportional to the number of devices $K$ as a result of aggregation. The rounds are repeated till the model converges. The commonly used convergence criteria require that the global gradient is sufficiently small and that loss function is evaluated to be below a given threshold, corresponding to reaching a target learning accuracy.

Another well-known framework is called \gls{pst}, which does not aim at leveraging mobile data, but instead attempts to harness computation resources distributed at many low-complexity devices for training a large-scale model. To this end, a server partitions the model into parts, called parameter blocks, and allocates each device one block for updating. \gls{pst} is based on the classic block coordinate descent algorithm, which is similar to \gls{sgd}. The \gls{pst} algorithm is similar to \gls{fl}, as described earlier, except for two differences. Firstly, the parameter server also downloads a training dataset from the cloud to devices in the first round. Secondly, each device is required to update only an assigned parameter block instead of the whole model. This reduces the computation complexity as well as the communication overhead of individual devices. 

\subsubsection{Statistical Characterization of Timing}
It is assumed that the server synchronizes the clocks of devices, and a time-reference point is defined by the server that initializes the learning process. The time spent on learning, called \emph{learning latency}, is measured from the instant when a server initiates the learning process to the instant when the global model converges. The  computation capacity of the edge cloud is much larger than each single device's, and the server's broadcasting latency is much shorter than the uploading latency of individual devices. The latency of each (communication) round is limited by the computation-plus-communication latency of devices. The information processing at a device is an example of the modularized architecture in Fig.~\ref{fig:CascadedModules}, which cascades the modules of local gradient/model computation, source encoding, and communication. Their efficiencies and latency performance  can be improved via joint design. Consider the joint design of computation and source encoding. For instance, the sparsity of a gradient/model can be exploited for achieving a large compression ratio (e.g., tens to hundreds times) without significant degradation of learning performance \cite{SongHanDeepGradientCompression}; given its geometry, a stochastic gradient is more suitably compressed  using a Grassmannian quantizer  instead of one using the \gls{mse} as the distortion measure \cite{YuqingGradientQuant}. Moreover, the rate information can be fed back from the communication module to control the gradient/model compression ratio. Unlike a point-to-point system, the effect of  gradient/model distortion due to a high compression ratio at  a particular device can be alleviated by update aggregation over many devices. The total latency of the cascaded on-device modules as well as those at the server in each communication round gives the per-round latency. 

Let $K$ denote the number of devices, $\Delta t_k^{(n)}$ the per-round latency of device $k$ in the $n$-th round, and $N$ the total number of rounds in the learning process. The aggregation operation in the \gls{fl} and \gls{pst} learning algorithms introduces an update synchronization constraint: in each round, the server needs to wait for all devices to finish their upload before the global model can be updated. Consequently, the learning latency can be written as 
\begin{equation}\label{LearnLatency}
\Delta t = \sum_{n=1}^N \max\left(\Delta t_1^{(n)}, \Delta t_2^{(n)}, \cdots, \Delta t_K^{(n)}\right). 
\end{equation}
Using the FL system in Fig.~\ref{fig:FL_Sys} as an example,  the three types of timing in distributed learning can be discussed as follows. 

\begin{itemize}
    \item \emph{Past Timing Reference}: In this case, considering a particular device, the usefulness of  a local-model update uploaded by the device depends on how much the current global model differs from the original one downloaded by the device (or, equivalently, its local model), where the downloading instant defines the time anchor. This can be translated into the number of updates to the global model performed by the server in the time in which the device computes  and transmits its local update. The case arises when the system comprises \emph{stragglers}. They refer to those devices that are slowest in computation or communication or both, which results in a latency bottleneck of the learning process. 
    One technique for coping with stragglers, termed \emph{lazy updating}, is for the server to only ask devices with large gradient norms, which indicate significant updates on the model, to upload their local gradient \cite{Giannakis:LazyUpdate:2018}. Another technique, called \emph{synchronous  updating}, is to reduce the stragglers' upload frequencies (i.e., uploading only once over multiple rounds)  while requiring other devices to perform an upload in each round \cite{AsynchronousFL:Xie:2019}. Such  techniques give rise to a trade-off between per-round latency and the required number of rounds. The lazier the straggler updates are, the lower the per-round latency is. On the other hand, this increases the staleness of the stragglers' updates and thereby causes the required number of rounds of the learning process to grow. Therefore, it is necessary to evaluate the staleness of stragglers' updates before they are applied to updating the global model. Moreover, to minimize  the learning latency, it is desirable to control the updating frequencies of individual stragglers depending on their computation and channel capacities. 
    
    \item \emph{Future Timing Reference}: It is often necessary to impose a deadline for distributed learning in a mobile network where the connections and donated computation resources of mobile devices, which are the data sources, are transient. For mission critical applications (e.g., disasters or robotic exploration of unknown environment), there is usually an urgency in acquiring the model. In addition, finite allocated radio resources may also limit the duration of channel use. In such cases, the needed duration for the learning task needs to be estimated. By placing a constraint on per-round latency $\Delta t$, the duration is measured by a planned  number of rounds, $N$, which is chosen to ensure the completion of the task with a high probability. To this end, we define the \emph{round-reliability} function as the probability that the model converges as
    \begin{equation}
        F_C(N) = \Pr(g(N, \bm{w}) \leq \delta \mid W, \mathcal{D}). 
    \end{equation}
    Then, given a target probability $p_0$, the required number of rounds is $N^\star = F_C(p_0)$. While the devices are assumed fixed in our exposition, it should be emphasized that in practice, they can vary from round to round due to mobility or scheduling. Though such randomness may incur some learning bias or data loss, there exist rich techniques to alleviate the effects, e.g., probabilistic or data-importance aware scheduling. 
    There exist several approaches for reducing the learning duration. One is to shorten the per-round communication latency by allocating more radio resources (power, bandwidth) or spatial degrees-of-freedom at the base station to support \gls{sdma} of more devices and enhance the spatial multiplexing of data. Researchers also design new communication techniques targeting federated learning such as over-the-air aggregation \cite{GX:AirAgg:2020}, scheduling \cite{Quek:ScheduleFL:2020,Song:ScheduleDL:2021},  and radio resource management \cite{Mingzhe:RRM:FL:2021}. Alternatively,  devices' computation speeds can be boosted  by increasing  their processors' clock frequencies  at the cost of higher power consumption \cite{Zeng:EnergyEfficientFL:ArXiv}. It is also possible to reduce the required number of rounds by selecting those devices with important data, i.e., data that are more informative for the model, to participate in model training \cite{Ren:ImportantAwareFL:2020}.  

    \item \emph{Relative Timing Reference}: The  reference moment $t=0$ refers to the instant the edge server initiates the learning process. The learning latency, denoted as $\Delta t$ in \eqref{LearnLatency},  measures the duration from $t=0$ until the instant when the learning criterion is met. In terms of the convergence criterion, extensive research has been conducted on quantifying the convergence speed, measured by the expectation of the averaged global gradient (or loss function) over rounds. The typical form of the speed is as follows \cite{bernstein2018signsgd,Gx:OneBitAirAgg:2021}:
    \begin{equation}
        \mathsf{E}\left[\frac{1}{N}\sum_{n=1}^N g(\bm{w}^{(n)})\right] \leqslant \frac{c_1\left\{ L(\bm{w}_0)-L(\bm{w}^\star) +\mathsf{E}\left[\frac{c_2}{K}\right]\right\}}{\sqrt{N}} 
        \label{Eq:dig_conv_single}
    \end{equation}
where $K$ is the number of connected devices, $\bm{w}_0$ the initial model, $\bm{w}^\star$ the optimal model, the descent step size is chosen as $\mu = \frac{c_3}{\sqrt{N}}$ and  $\{c_1, c_2, c_3\}$ are constants. Note that $K$ is a random variable due to fading in wireless links and the constants $\{c_1, c_2\}$ depends on wireless parameters such as signal-to-noise ratios and outage probabilities \cite{Gx:OneBitAirAgg:2021}. The convergence result allows the estimation of the required number of rounds. The mentioned techniques on reducing per-round latency are also applicable in this case. 

\end{itemize}

\subsection{Timing in Edge Inference} \label{sec:edgeinference}
The preceding subsection focuses on the training of machine learning algorithms. The theme of this subsection is  their application, called edge inference. Specifically, edge inference refers to making intelligent predictions and decisions at an edge server based on data generated by \gls{iot} devices,  which finds practically unlimited applications, ranging from smart cities to autonomous driving to smart wearable. Compared with on-device inference, edge inference has the advantages of operating a large-scale \gls{ai} model (e.g., the Google-Cloud classifier that can render 700 image classes), enabling centralized decision and control in an \gls{iot} scenario with many sensors, and continuous \gls{ai} model improvements using aggregated data or distributed learning. 

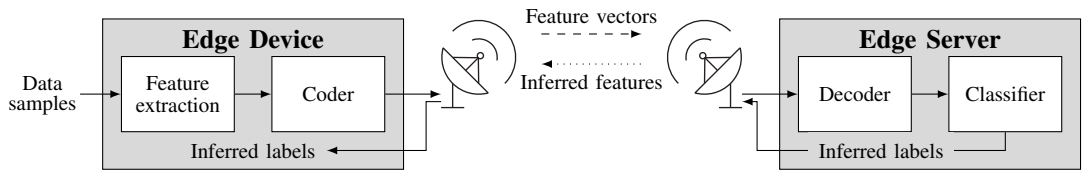
\begin{figure}[t]
    \centering
    \input{figures/edge_update}\vspace{0.3cm}
    \caption{Edge inference system and its operations.}
    \label{fig:EdgeInference_Sys}
\end{figure}

To make the discussion concrete, consider the edge learning system in Fig.~\ref{fig:EdgeInference_Sys}.
The information processing in the system  is  another  example of the modularized architecture in Fig.~\ref{fig:CascadedModules}, which cascades the on-device modules including feature extraction (computation), source encoder, and transmitter and   the server modules including the receiver, source decoder, and classifier (computation). Their operations, timing, and trade-offs are discussed as follows. Usually, only a small fraction of information embedded in a high-dimensional sample of a raw-data distribution, denoted as a $N\times 1$ vector $\mathbf{d}$,  is useful for inference. To reduce the communication overhead, the sample is compressed by projection onto a low-dimensional space that contains the most important information of the data distribution, called a \emph{feature space}. The operation is called \emph{feature extraction}, and the result is called an \emph{feature vector}, denoted as a $M\times 1$ vector $\mathbf{x}$ with $M \ll N$. A classic technique for feature extraction, called \gls{pca}, computes the feature space as a  \emph{linear subspace} \cite{PCATutorial:Abdi:2010}. The modern approach applies a neural network, called \emph{encoder},  to identify a \emph{nonlinear} feature space, which yields better inference performance than \gls{pca}, at the cost of higher complexity \cite{JunZhang:EdgeInference:2020}. The feature vector (or part of it) is then transmitted reliably to the server as regular data (i.e., by digital modulation/demodulation and coding/decoding) and fed into another neural network, called {decoder}, for generating  the predicted value.  Alternatively, the wireless channel can be treated as a part of the encoder and trained jointly with the decoder to achieve satisfactory inference performance in the presence of channel distortion \cite{Gunduz:ImageRetrieval:2021}. Due to imperfect sensing conditions, there is always uncertainty in prediction, which can be measured  using different metrics, such as entropy. Let $U(\mathbf{y})$ denote the uncertainty of prediction on the received  feature vector, $\mathbf{y}$, containing a subset of the features in $\mathbf{x}$. $U(\mathbf{y})$ is a monotone decreasing function of the feature subset. As low uncertainty can be  usually translated into high prediction accuracy, from the inference perspective, it is desirable for the device  to transmit as many features as possible. However, the transmitted features may have to be limited due to a constraint on radio resources or latency. Given this trade-off, we can discuss timing in edge inference as follows.

\begin{itemize}
    \item \emph{Past Timing Reference}: The basic operation of an \gls{iot} system is to aggregate data from a large number of sensors, making inference and decisions, and then transmit commands to actuators for execution. Many \gls{iot} applications, such as vehicle-accident avoidance, crime detection and prevention, and smart manufacturing, are latency sensitive. Thus, the value of sensing data decreases with their age. On the other hand, the heterogeneity in computation capabilities, locations, and link reliability and bandwidth can cause the data (or features) transmitted by sensors to arrive at servers with different ages. To ensure the accuracy of inference and quality of decisions requires the server to select uploaded data as inference inputs by considering their staleness. This gives rise to a trade-off between the age of inference output and data diversity, affecting the inference accuracy. For instance, the prediction of a traffic accident is less accurate given sensing data from fewer nearby vehicles, and the recognition of an object/human-being is more accurate with multiple camera observations from different perspectives. Such trade-offs can be optimized by scheduling, radio resource management, and evaluation of the importance of sensing data (e.g., a rare event or a sample in a minority class of imbalanced data). 
    
    \item \emph{Future Timing Reference}: Many timing-sensitive applications, such as autonomous driving and \gls{vr},  require a device to receive the inference result within a fixed time duration (usually ranging from tens to hundreds of milliseconds) from the instant of sensing. The required fast responses  are essential for an auto-pilot to prevent accidents or a \gls{vr} device to avoid causing dizziness to the user. This limits the cumulative timing for sensing, round-trip communication, and overall computation. Given a deadline or under a constraint on inference accuracy, the latency can be reduced by using a simple technique for on-device feature extraction (e.g., \gls{pca}) that is compensated for by deploying a complex high-performance deep neural network for classification at the server, transmitting only the minimum number of features and allocating  sufficient radio resources for  transmission. 
    
    \item \emph{Relative Timing Reference}: Mobile devices are usually constrained in computation and communication resources. Given a target inference accuracy, it is desirable to extract and transmit only the minimum number of features. However, the number cannot be estimated in advance due to the geographic separation of data and \gls{ai} model. The problem can be solved using a  progress feature transmission protocol. The essential idea is to transmit the features block by block until the server confirms that the desired accuracy is met. The communication latency for edge inference on a sample is stochastic, as it depends on the sample and channel realizations and the data distribution, among other factors. Designing the protocol requires the use of an accuracy measure such as the uncertainty function $U$ discussed earlier, as the uncertainty computation at the server uses a deep neural network model and its feedback to the device. For data containing objects with weak differentiability, reaching the target accuracy may require the transmission of a large number of features or even fail when all features are transmitted. In such cases, to avoid excessive communication  overhead, the device need to predict and balance the communication cost and uncertainty reduction from additional feature transmission and decide on when to stop transmission. The policy design can be formulated as an optimal stopping problem \cite{Chow71,Shiryaev78}. 
\end{itemize}

\section{Conclusion and Future Vision}\label{sec:concl}
Wireless connectivity is the cornerstone of the digital technologies that bridge the gap between the physical and digital world. Wireless connections offer remote interaction among humans and machines over extended distances. This calls for a careful system optimization to conform to the measurement and perception of time in the physical and digital realm. 

5G is a first step towards providing the ultimate connectivity, i.e., a communication system that can flexibly support any time of timing requirement. We can think of the three connectivity types (\gls{embb}, \gls{mmtc}, and \gls{urllc}) defined in it as different axes in a wider \emph{service space}, which includes multiple kinds of performance. Services can then be seen as combinations of traffic flows belonging to one of the three services: for example, we can imagine a manufacturing scenario in which sensors and \gls{iot} devices gather information about the environment (\gls{mmtc}), while robots and drones send a video or depth map feed to navigate in the environment (\gls{embb}) and give a central controller information about future actions, which are implemented by reliably transmitting real-time commands (\gls{urllc}) to the same robots and drones. In this sense, the approach of 5G technology to satisfy certain end-to-end real-time requirements has been maximalistic: service requirements are defined only in the \gls{ran}, with very strict timing requirements (e.g., \gls{urllc} allocates resources so that the wireless transmission consumes a very small, predictable part of the overall timing budget), in order to allow the maximum possible flexibility to other components of the system.

However, as wireless systems evolve towards 6G, the ambition to immerse the digital into the physical reality will increase and the real-time requirements posed to the wireless connectivity will become even more stringent. Recent research has brought a number of other timing measures, such as the Age of Information or Age of Loop , that are more suitable to characterize the overall real-time operation compared to a mere optimization of a latency parameter. A more general framework of the operations that involve the network is then necessary, as well as a wider view of timing: in this paper we have provided a panoramic view on the field of timing measures and defined a general statistical framework that offers their systematic characterization. 

In our vision, the interaction between the digital and physical world is a key component of future networks: as the constraints on timing are often inherently analog, i.e., depend on decisions and events in the physical world, the digital part of the system, consisting of communication, computation, memory, and control, needs to be considered as a whole, and timing optimization needs to become more flexible than traditional latency minimization. The development of metrics such as \gls{aoi} and \gls{voi}, as well as the integration of applications such as learning and consensus protocols into the timing characterization, is
a step towards a more holistic view of communication, in which the most important data are transmitted at the right time to enable the applications, associating \emph{semantics} to measurements and signaling~\cite{popovski2020semantic,kountouris2020semantics,uysal2021semantics}. This transformation is critical for providing a new type of connectivity, which can go beyond the limited, predefined classes of 5G.

In this work, we have presented the immediate and future usefulness of the proposed framework in different communication models, identified the basic trade-offs and established the relation between different types of characterization of timing. The objective of the paper is to provide a tutorial view on this emerging area and offer the general statistical framework for timing as a tool for defining and solving problems in real-time wireless communication systems.

\bibliographystyle{IEEEtran}
\bibliography{TimingReferences}

\end{document}

%% file: figures/basic_scenario.tex
\begin{tikzpicture}[>=latex]

\node[inner sep=0pt] (bs) at (0,0) {\includegraphics[width=.2\textwidth]{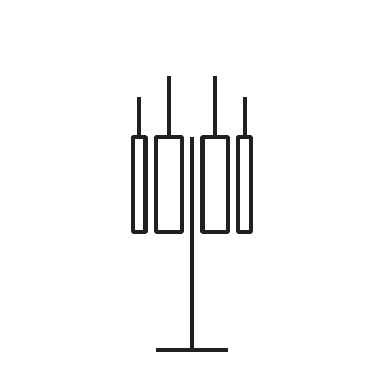}};
\node[inner sep=0pt] (usr) at (-4,-2) {\includegraphics[width=.1\textwidth]{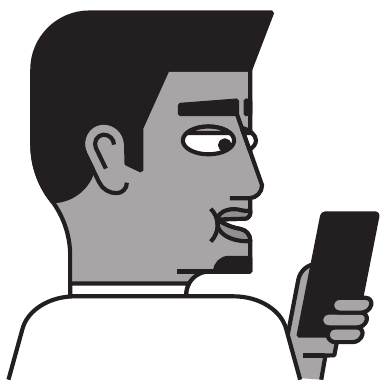}};
\node[inner sep=0pt] (exc) at (4,-1.5) {\includegraphics[width=.2\textwidth]{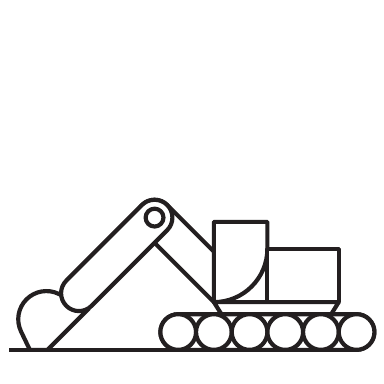}};

\draw[->,color=green,line width=1pt] (usr.east) -- node [below,midway,sloped] {\vspace{0.1cm}\textcolor{black}{\scriptsize{High-rate video}}} (-0.5,-1);
\draw[->,color=red,line width=1pt] (3,-1.7) -- node [below,midway,sloped] {\vspace{0.1cm}\textcolor{black}{\scriptsize{Intermittent updates}}} (0.5,-1);

\node (a) at (0.25, -3.75) {(a)};
\foreach \i in {1,2,3,5,6,7,9,10,11,13,14,15,17,18,19}
    \foreach \j in {1,...,3}
        \draw[fill=white!50!green,draw=none] (\i/2-5.25,-4) rectangle (\i/2-4.75, -4.5);;
\foreach \i in {4,8,12,16,20}
    \draw[fill=white!50!red,draw=none] (\i/2-5.25,-4) rectangle (\i/2-4.75, -4.5);

\draw[->,line width=0.4pt] (-4.75,-4.6) -- node [below,pos=0.95] {\vspace{0.1cm}\textcolor{black}{\scriptsize{Time}}} (5.25,-4.6);
\draw[->,line width=0.4pt] (-4.75,-6.1) -- node [below,pos=0.95] {\vspace{0.1cm}\textcolor{black}{\scriptsize{Time}}} (5.25,-6.1);

\node (b) at (0.25, -5.25) {(b)};
\foreach \i in {1,2,3,5,6,7,8,9,10,11,12,13,14,16,17,18,19,20}
    \foreach \j in {1,...,3}
        \draw[fill=white!50!green,draw=none] (\i/2-5.25,-6) rectangle (\i/2-4.75, -5.5);;
\foreach \i in {4,12,15}
    \draw[fill=white!50!red,draw=none] (\i/2-5.25,-6) rectangle (\i/2-4.75, -5.5);

\vspace{0.3cm}

\end{tikzpicture}

%% file: figures/timing_model.tex
\tikzset{  
block/.style    = {draw, rectangle, minimum height = 2cm, minimum width = 1em}}
\begin{tikzpicture}[auto]

\node[name=app1, draw, orange, minimum height=1.5cm, minimum width=2cm] at (-0.5,5) {App};
\node[name=hl1, draw, orange, minimum height=1.5cm, minimum width=2cm] at (2.5,5) {HL};
\node[name=ll1, draw, cyan, minimum height=1.5cm, minimum width=2cm] at (5,5) {LL};
\node[name=ch, draw, cyan, minimum height=1.5cm, minimum width=3cm] at (8,5) {Channel};
\node[name=chdesc,text width=2.75cm,align=center] at (8,6.5) {5G comm. service};
\node[name=ll2, draw, cyan, minimum height=1.5cm, minimum width=2cm] at (11.5,5) {LL};
\node[name=hl2, draw, orange, minimum height=1.5cm, minimum width=2cm] at (13.5,5) {HL};
\node[name=app2, draw, orange, minimum height=1.5cm, minimum width=2cm] at (16,5) {App};

\draw[dashed] (1.25,6) to (1.25,-2);
\draw[dashed] (3.75,6) to (3.75,-2);
\draw[dashed] (6.25,6) to (6.25,-2);
\draw[dashed] (9.75,6) to (9.75,-2);
\draw[dashed] (12.25,6) to (12.25,-2);
\draw[dashed] (14.75,6) to (14.75,-2);

\draw[->] (app1.east) to (hl1.west);
\draw[->] (hl1.east) to (ll1.west);
\draw[->] (ll1.east) to (ch.west);
\draw[->] (ch.east) to (ll2.west);
\draw[->] (ll2.east) to (hl2.west);
\draw[->] (hl2.east) to (app2.west);

\draw[->] (0.5,4) to (1.25,4);
\draw[->] (0.5,2) to (1.25,2);
\draw[->] (0.5,0.25) to (1.25,0.25);

\draw[->] (14.75,0.6) to (15.5,0.6);
\draw[->] (14.75,-1.3) to (15.5,-1.3);

\draw[|-|,orange] (0.25,4) to node[midway,left,text width=1.75cm,align=center,font={\small}] {App transfer interval} (0.25,2);
\draw[|-|,cyan] (3.5,3.25) to node[midway,left,text width=1.75cm,align=center,font={\small}] {5G transfer interval} (3.5,1);
\draw[|-|,cyan] (12.5,3.25) to node[midway,right,text width=1.75cm,align=center,font={\small}] {End-to-end latency} (12.5,1.5);
\draw[|-|,orange] (15,4) to node[midway,right,text width=1.75cm,align=center,font={\small}] {Transmission time} (15,0.75);
\draw[|-|,orange] (15,0.6) to node[midway,right,text width=1.75cm,align=center,font={\small}] {Update time} (15,-1.3);

\draw[->,orange] (1.25,3.75) to (3.75,3.25);
\draw[->,cyan] (3.75,3.2) to (6.25,3);
\draw[->,cyan] (6.25,2.8) to (9.75,2.1);
\draw[->,cyan] (9.75,2) to (12.25,1.5);
\draw[->,orange] (12.25,1.25) to (14.75,0.75);

\draw[->,orange] (1.25,1.75) to (3.75,1);
\draw[->,cyan] (3.75,0.9) to (6.25,0.6);
\draw[->,cyan] (6.25,0.45) to (9.75,-0.1);
\draw[->,cyan] (9.75,-0.3) to (12.25,-0.6);
\draw[->,orange] (12.25,-0.7) to (14.75,-1.2);

\draw[->,orange] (1.25,0) to (3.75,-0.6);
\draw[->,cyan] (3.75,-0.8) to (6.25,-1.1);
\draw[->,cyan] (6.25,-1.35) to (9.75,-1.8);
\draw[->,cyan,dashed] (9.75,-1.9) to (11,-2);

\draw[densely dotted,orange] (1.25,4) to (14.75,4);
\draw[densely dotted,cyan] (3.75,3.25) to (12.25,3.25);

\end{tikzpicture}

%% file: figures/oran.tex
\begin{tikzpicture}[>=latex]

\draw (0,5) -- (1,5) -- (0.5,0) -- cycle;
\fill[blue!20!white](0,5) -- (1,5) -- (0.5,0) -- cycle;
\node (txt1) at (0.5,5.5) {\scriptsize{Latency for}};
\node (txt2) at (0.5,5.2) {\scriptsize{control}};

\node[rectangle,draw,fill=blue!60!white,minimum width=1.5cm,minimum height=0.5cm] (round1) at (0.5,4.5) {\textcolor{white}{\scriptsize{$\gg1$ s}}};
\node[rectangle,draw,fill=blue!60!white,minimum width=1.5cm,minimum height=0.5cm] (round2) at (0.5,3.2) {\textcolor{white}{\scriptsize{$10$-$100$ ms}}};
\node[rectangle,draw,fill=blue!60!white,minimum width=1.5cm,minimum height=0.5cm] (round3) at (0.5,1.5) {\textcolor{white}{\scriptsize{Real-time}}};

\node[rectangle,fill=gray!40!white,minimum width=6cm,minimum height=0.5cm] (box1) at (5,4.5) {\scriptsize{Wireless domain management}};
\node[ellipse,fill=green!40!white,minimum width=0.6cm,minimum height=0.4cm] (ai1) at (7.65,4.5) {};
\node at (7.65,4.5){\scriptsize{AI}};

\node[rectangle,fill=blue!60!white,minimum width=6cm,minimum height=1.5cm] (box2) at (5,3.2) {};
\node[ellipse,fill=green!40!white,minimum width=0.6cm,minimum height=0.4cm] (ai2) at (7.65,3.7) {};
\node at(5,3.75) {\textcolor{white}{\scriptsize{RAN Intelligent Controller (RIC)}}};
\node at (7.65,3.7){\scriptsize{AI}};
\node[rectangle,fill=gray!40!white,minimum width=5cm,minimum height=0.4cm] (bigbox1) at (5,2.75) {\tiny{RAN data analytics \& AI platform}};
\node[rectangle,fill=gray!60!white,minimum width=1.1cm,minimum height=0.55cm] (smallbox1) at (3.2,3.35) {};
\node[rectangle,fill=gray!60!white,minimum width=1.1cm,minimum height=0.55cm] (smallbox2) at (4.4,3.35) {};
\node[rectangle,fill=gray!60!white,minimum width=1.1cm,minimum height=0.55cm] (smallbox3) at (5.6,3.35) {};
\node[rectangle,fill=gray!60!white,minimum width=1.1cm,minimum height=0.55cm] (smallbox4) at (6.8,3.35) {};
\node at (3.2,3.5){\tiny{QoS}};
\node at (3.2,3.2){\tiny{optimization}};
\node at (4.4,3.5){\tiny{Slicing}};
\node at (4.4,3.2){\tiny{optimization}};
\node at (5.6,3.5){\tiny{Mobility}};
\node at (5.6,3.2){\tiny{optimization}};
\node at (6.8,3.5){\tiny{Third party}};
\node at (6.8,3.2){\tiny{application}};

\node[rectangle,fill=blue!20!white,minimum width=6cm,minimum height=1.2cm] (box3) at (5,1.5) {};
\node[rectangle,fill=gray!20!white,minimum width=1.2cm,minimum height=0.5cm] (subbox1) at (4.3,1.8) {\scriptsize{CU-CP}};
\node[rectangle,fill=gray!20!white,minimum width=1.2cm,minimum height=0.5cm] (subbox2) at (7,1.8) {\scriptsize{CU-UP}};
\node[rectangle,fill=gray!20!white,minimum width=1.2cm,minimum height=0.4cm] (subbox3) at (4.3,1.2) {\scriptsize{DU}};
\node[rectangle,fill=gray!20!white,minimum width=1.2cm,minimum height=0.4cm] (subbox4) at (7,1.2) {\scriptsize{RRU}};
\draw (subbox1.east) -- node[midway,above]{\tiny{3GPP E1}} (subbox2.west);
\draw (subbox3.east) -- node[midway,above]{\tiny{ORAN NGF-I}} (subbox4.west);
\node at(2.5,1.8) {\small{RAN}};

\draw (box1.south) -- node[midway,right]{\scriptsize{A1}} (box2.north);
\draw (box2.south) -- node[midway,right]{\scriptsize{E2}} (box3.north);

\draw (box1.east) -- node[midway,above]{\scriptsize{O1}}(8.5,4.5) -- (8.5,1.5) -- (box3.east);
\draw (box2.east) -- (8.5,3.2);

\node[circle,draw,green!80!white,fill=white,minimum width=0.6cm] at (-0.3,4.8) {};

  \draw[green!80!white,fill=green!20!white]  (-0.3,5.06) arc (90:300:2.6mm) --
  (-0.3,4.8)
  -- cycle;

  \draw[green!80!white] (-0.26,5.1) -- (-0.27,5.15) -- (-0.2,5.15) -- (-0.2,5.15) arc (-90:90:0.4mm) -- (-0.4,5.23) -- (-0.4,5.23) arc (90:270:0.4mm) -- (-0.33,5.15) -- (-0.34, 5.1);

\node[circle,draw,green!80!white,fill=white,minimum width=0.5cm] at (-0.3,3.5) {};

  \draw[green!80!white,fill=green!20!white]  (-0.3,3.71) arc (90:300:2.1mm) --
  (-0.3,3.5)
  -- cycle;

  \draw[green!80!white] (-0.265,3.75) -- (-0.275,3.8) -- (-0.22,3.8) -- (-0.22,3.8) arc (-90:90:0.3mm) -- (-0.38,3.86) -- (-0.38,3.86) arc (90:270:0.3mm) -- (-0.325,3.8) -- (-0.335, 3.75);

\node[circle,draw,green!80!white,fill=white,minimum width=0.4cm] at (-0.3,1.8) {};

  \draw[green!80!white,fill=green!20!white]  (-0.3,1.96) arc (90:300:1.6mm) --
  (-0.3,1.8)
  -- cycle;

  \draw[green!80!white] (-0.27,2) -- (-0.28,2.05) -- (-0.23,2.05) -- (-0.23,2.04) arc (-90:90:0.2mm) -- (-0.37,2.08) -- (-0.37,2.08) arc (90:270:0.2mm) -- (-0.32,2.04) -- (-0.33, 2);

\end{tikzpicture}

%% file: figures/references.tex
\begin{tikzpicture}[>=latex]

\draw[-latex] (0,3) --  (10,3);
\draw[-latex] (0,0) --  (10,0);
\node at (9.7,-0.25){\small{Time}};
\node at (9.7,2.75){\small{Time}};
\node at (-0.8,0.15){\scriptsize{Node 2: edge}};
\node at (-0.8,-0.15){\scriptsize{controller}};
\node at (-0.6,3.15){\scriptsize{Node 1:}};
\node at (-0.6,2.85){\scriptsize{sensor}};

\draw[cyan,arrows={<->[scale=1,cyan]}] (8.5,3.5) -- node[midway,below]{\textcolor{cyan}{\scriptsize{$T_s$}}} (9.5,3.5);
\draw[cyan] (9.5,3.4) -- (9.5,3.6);
\draw[cyan] (8.5,3.4) -- (8.5,3.6);

\draw[-latex] (2,3) --  (2.7,0);
\draw[-latex] (2.7,0) -- node[midway,above,sloped]{\scriptsize{ACK}} (3.5,3);
\draw[-latex] (7.7,0) --  (8.5,3);
\draw[-latex] (8.5,3) -- node[midway,above,sloped]{\scriptsize{ACK}} (9,0);
\draw[-latex] (4.3,3) --  (5.1,0);
\draw[-latex] (5.7,3) --  (6.1,0);
\draw[-latex] (6.1,0) -- node[midway,above,sloped]{\scriptsize{ACK}} (6.5,3);
\draw (2,2.9) -- (2,3.1);
\draw (3.5,2.9) -- (3.5,3.1);
\draw (3.8,2.9) -- (3.8,3.1);
\draw (4.3,2.9) -- (4.3,3.1);
\draw (5.7,2.9) -- (5.7,3.1);
\draw (8.5,2.9) -- (8.5,3.1);
\draw (6.5,2.9) -- (6.5,3.1);
\draw (2.7,-0.1) -- (2.7,0.1);
\draw (7.7,-0.1) -- (7.7,0.1);
\draw (9,-0.1) -- (9,0.1);
\draw (5.1,-0.1) -- (5.1,0.1);
\draw (6.1,-0.1) -- (6.1,0.1);

\draw (3.8,3) -- (4.2,1.5);
\draw[red,line width=1pt] (4.1,1.6) -- (4.3,1.4);
\draw[red,line width=1pt] (4.3,1.6) -- (4.1,1.4);
\draw (5.1,0) -- (5.4,1.5);
\draw[red,line width=1pt] (5.5,1.6) -- (5.3,1.4);
\draw[red,line width=1pt] (5.3,1.6) -- (5.5,1.4);

\draw[cyan,arrows={<->[scale=1,cyan]}] (2.7,-0.5) -- node[midway,below]{\textcolor{cyan}{\scriptsize{$T_d$}}} (7,-0.5);
\draw[cyan] (2.7,-0.4) -- (2.7,-0.6);
\draw[cyan] (7,-0.4) -- (7,-0.6);

\draw[green] (2,3) -- (2,4.2);
\draw[green] (3.8,3) -- (3.8,3.6);
\draw[green] (4.3,3) -- (4.3,4.1);
\draw[green] (5.7,3) -- (5.7,4.7);
\node at (2,4.2)[circle,green,fill=green,inner sep=1pt]{};
\node at (3.8,3.6)[circle,green,fill=green,inner sep=1pt]{};
\node at (4.3,4.1)[circle,green,fill=green,inner sep=1pt]{};
\node at (5.7,4.8)[circle,green,fill=green,inner sep=1pt]{};

\node at (4.3,4.3){\textcolor{green}{\scriptsize{$s_5$}}};
\node at (5.7,5){\textcolor{green}{\scriptsize{$s_6$}}};
\node at (3.8,3.8){\textcolor{green}{\scriptsize{$s_4$}}};
\node at (2,4.4){\textcolor{green}{\scriptsize{$s_1$}}};

\node at (2.15,3.2){\scriptsize{$t_1$}};
\node at (2.7,-0.2){\scriptsize{$t_2$}};
\node at (3.5,3.2){\scriptsize{$t_3$}};
\node at (3.95,3.2){\scriptsize{$t_4$}};
\node at (4.45,3.2){\scriptsize{$t_5$}};
\node at (5.85,3.2){\scriptsize{$t_6$}};
\node at (6.5,3.2){\scriptsize{$t_8$}};
\node at (6.1,-0.2){\scriptsize{$t_7$}};
\node at (7.55,0.2){\scriptsize{$t_9$}};
\node at (7.7,-0.2){\scriptsize{$c_1$}};
\node at (8.5,3.2){\scriptsize{$t_{10}$}};
\node at (9,-0.2){\scriptsize{$t_{11}$}};

\draw plot[smooth] coordinates {(1,4.4) (1.2,4.5) (1.3,4.7)(1.5,4.6)(1.8,4.1)(2,4.2)(2.3,4.4)(2.6,4.1)(2.8,4.4)(3.1,4.3)(3.4,4)(3.5,3.85)(3.7,3.65)(3.8,3.6)(4,3.7)(4.3,4.1)(4.5,4.2)(4.7,4.2)(5,4.4)(5.3,4.7)(5.7,4.8)(6.1,4.7)(6.6,4.1)(6.9,3.9)(7.2,3.8)(7.3,3.8)(7.5,4)(7.8,4.1)(8,4.3)(8.2,4.4)(8.3,4.4)(8.6,4.2)(8.8,4.1)(8.9,4.1)(9.1,4.3)(9.3,4.3)(9.5,4)};

\end{tikzpicture}

%% file: figures/layer_model.tex
\tikzset{  
block/.style    = {draw, rectangle, minimum height = 2cm, minimum width = 1em}}
\begin{tikzpicture}[auto]

\node[name=app] at (-6,1.5) {\large \textbf{Application}};
\node[name=hl] at (-6,-1.5) {\large \textbf{Higher layer}};
\node[name=ll] at (-6,-4.5) {\large \textbf{Lower layer}};
\node[name=src] at (-1,4) {\large \textbf{Source}};
\node[name=dst] at (10,4) {\large \textbf{Destination}};
\draw[dashdotted] (-8,0) to (14.5,0);
\draw[dashdotted] (-8,-3) to (14.5,-3);
\draw[dashdotted] (-8,-3) to (14.5,-3);

\node[name=proc, draw, orange, minimum height=1.5cm, minimum width=2.5cm] at (0,1.5) {\large Process};
\node[name=samp, draw, orange, minimum height=1.5cm, minimum width=2.5cm] at (3, 1.5) {\large Sampling};
\node[name=mon, draw, orange, minimum height=1.5cm, minimum width=2.5cm] at (12,1.5) {\large Monitor};
\node[name=est, draw, orange, minimum height=1.5cm, minimum width=2.5cm] at (9,1.5) {\large Estimator};

\node[name=fb, draw, green, minimum height=1.5cm, minimum width=2.5cm] at (9,-1.5) {\large Feedback};
\node[name=ch, draw, green, minimum height=1.5cm, minimum width=2.5cm] at (3,-1.5) {\large Scheduling};

\node[name=tx, draw, cyan, minimum height=1.5cm, minimum width=2.5cm] at (3,-4.5) {\large Transceiver};
\node[name=rx, draw, cyan, minimum height=1.5cm, minimum width=2.5cm] at (7,-4.5) {\large Transceiver};

\draw[-Latex,orange] (proc.east) to (samp.west);
\draw[-Latex,orange] (est.east) to (mon.west);

\draw[-,cyan] (fb.south) to node[near end,right,font={\large}] {Feedback} (9,-4.2);
\draw[-Latex,cyan] (9,-4.2) to ([yshift=0.3cm]rx.east);
\draw[-Latex,green] (est.south) to node[near end,right,font={\large}] {Estimate} (fb.north);
\draw[-Latex,green] (samp.south) to node[near end,left,font={\large}] {Sampled data} (ch.north);
\draw[-Latex,cyan] ([xshift=0.3cm]ch.south) to node[near end,right,font={\large}] {Data} ([xshift=0.3cm]tx.north);
\draw[Latex-,green] ([xshift=-0.3cm]ch.south) to node[near end,left,font={\large}] {Feedback} ([xshift=-0.3cm]tx.north);
\draw[Latex-Latex,cyan] (tx.east) to (rx.west);
\draw[orange] (rx.north) to node[near end,left,font={\large}] {Data} (7,1.5);
\draw[-Latex,orange] (7,1.5) to (est.west);

\draw (5,4.5) to (5,-6.5);
\draw (-4.5,4.5) to (-4.5,-6.5);

\end{tikzpicture}

%% file: figures/layer_model_closedloop.tex
\tikzset{  
block/.style    = {draw, rectangle, minimum height = 2cm, minimum width = 1em}}
\begin{tikzpicture}[auto]

\node[name=app] at (-6,1.5) {\large \textbf{Application}};
\node[name=hl] at (-6,-1.5) {\large \textbf{Higher layer}};
\node[name=ll] at (-6,-4.5) {\large \textbf{Lower layer}};
\node[name=src] at (-1,4) {\large \textbf{Source}};
\node[name=dst] at (10,4) {\large \textbf{Destination}};
\draw[dashdotted] (-8,0) to (14.5,0);
\draw[dashdotted] (-8,-3) to (14.5,-3);
\draw[dashdotted] (-8,-3) to (14.5,-3);

\node[name=proc, draw, orange, minimum height=1.5cm, minimum width=2.5cm] at (0,1.5) {\large Process};
\node[name=samp, draw, orange, minimum height=1.5cm, minimum width=2.5cm] at (3, 1.5) {\large Sampling};
\node[name=mon, draw, orange, minimum height=1.5cm, minimum width=2.5cm] at (12,1.5) {\large Control};
\node[name=est, draw, orange, minimum height=1.5cm, minimum width=2.5cm] at (9,1.5) {\large Estimator};

\node[name=fb, draw, green, minimum height=1.5cm, minimum width=2.5cm] at (9,-1.5) {\large Feedback};
\node[name=ch, draw, green, minimum height=1.5cm, minimum width=2.5cm] at (3,-1.5) {\large Scheduling};

\node[name=tx, draw, cyan, minimum height=1.5cm, minimum width=2.5cm] at (3,-4.5) {\large Transceiver};
\node[name=rx, draw, cyan, minimum height=1.5cm, minimum width=2.5cm] at (7,-4.5) {\large Transceiver};

\draw[-,cyan] (fb.south) to node[near end,right,font={\large}] {Feedback} (9,-4.2);
\draw[-Latex,cyan] (9,-4.2) to ([yshift=0.3cm]rx.east);
\draw[-Latex,green] (est.south) to node[near end,right,font={\large}] {Estimate} (fb.north);
\draw[-Latex,green] (samp.south) to node[near end,left,font={\large}] {Sampled data} (ch.north);
\draw[-Latex,cyan] ([xshift=0.3cm]ch.south) to node[near end,right,font={\large}] {Data} ([xshift=0.3cm]tx.north);
\draw[Latex-,green] ([xshift=-0.3cm]ch.south) to node[near end,left,font={\large}] {Feedback} ([xshift=-0.3cm]tx.north);
\draw[Latex-Latex,cyan] (tx.east) to (rx.west);
\draw[orange] (rx.north) to node[near end,left,font={\large}] {Data} (7,1.5);
\draw[-Latex,orange] (7,1.5) to (est.west);

\draw (5,4.5) to (5,-6.5);
\draw (-4.5,4.5) to (-4.5,-6.5);

\node[name=act, draw, orange, minimum height=1.5cm, minimum width=2.5cm] at (-3,1.5) {\large Actuator};

\draw[cyan] (mon.south) to node[midway,right,font={\large}] {Command} (12,-4.8);
\draw[-Latex,cyan] (12,-4.8) to ([yshift=-0.3cm]rx.east);

\draw[Latex-,orange] (act.south) to node[midway,right,font={\large}] {Command} (-3,-4.5);
\draw[orange] (-3,-4.5) to (tx.west);

\end{tikzpicture}

%% file: figures/shannon.tex
\begin{tikzpicture}

\node[rectangle,draw,dashed,line width=0.4pt,minimum width=2.5cm,minimum height=5cm] (node1) at (0,0) {};
\node[rectangle,draw,dashed,line width=0.4pt,minimum width=2.5cm,minimum height=5cm] (node2) at (5,0) {};

\node (txt1) at (0,2.25) {Node 1};
\node (txt2) at (5,2.25) {Node 2};

\node[rectangle,draw,line width=1pt,minimum width=1.5cm,minimum height=1cm] (HTR1) at (0,1) {};
\node[rectangle,draw,line width=1pt,minimum width=1.5cm,minimum height=1cm] (LTR1) at (0,-1.5) {\scriptsize{Transmitter}};
\node[rectangle,draw,line width=1pt,minimum width=1.5cm,minimum height=1cm] (HTR2) at (5,1) {\scriptsize{Destination}};
\node[rectangle,draw,line width=1pt,minimum width=1.5cm,minimum height=1cm] (LTR2) at (5,-1.5) {\scriptsize{Receiver}};
\node[rectangle,draw,line width=1pt,minimum width=1.5cm,minimum height=1cm] (ch) at (2.5,-1.5) {\small{$p(y|x)$}};

\node (h1) at (0,1.25) {\scriptsize{Information}};
\node (h2) at (0,0.75) {\scriptsize{source}};

\draw[-latex] (HTR1.south) --  node [left,near end] {\scriptsize{Message}}(LTR1.north);
\draw[latex-] (HTR2.south) -- node [right,midway,text width=1.5cm,font=\scriptsize\linespread{0.8}] { Decoded message}(LTR2.north);
\draw[-latex] (ch.east) -- node [above,near end] {\small{$y$}} (LTR2.west);
\draw[latex-] (ch.west) -- node [above,near end] {\small{$x$}}(LTR1.east);

\draw[-,dashed,line width=1.2pt] (-5,-0.25) -- (7,-0.25);

\node (hi) at (-3,0.1) {\small{High Layer (HL)}};
\node (lo) at (-3,-0.7) {\small{Low Layer (LL)}};

\end{tikzpicture}

%% file: figures/twoway.tex
\begin{tikzpicture}

\node[rectangle,draw,dashed,line width=0.4pt,minimum width=2.5cm,minimum height=5cm] (node1) at (0,0) {};
\node[rectangle,draw,dashed,line width=0.4pt,minimum width=2.5cm,minimum height=5cm] (node2) at (5,0) {};

\node (txt1) at (0,2.25) {Node 1};
\node (txt2) at (5,2.25) {Node 2};

\node[rectangle,draw,line width=1pt,minimum width=1.5cm,minimum height=1cm] (HTR1) at (0,1) {\small{HTR1}};
\node[rectangle,draw,line width=1pt,minimum width=1.5cm,minimum height=1cm] (LTR1) at (0,-1.5) {\small{LTR1}};
\node[rectangle,draw,line width=1pt,minimum width=1.5cm,minimum height=1cm] (HTR2) at (5,1) {\small{LTR1}};
\node[rectangle,draw,line width=1pt,minimum width=1.5cm,minimum height=1cm] (LTR2) at (5,-1.5) {\small{LTR2}};
\node[rectangle,draw,line width=1pt,minimum width=1.5cm,minimum height=1cm] (ch) at (2.5,-1.5) {};

\node (ch1) at (2.5,-1.25) {\small{Two-way}};
\node (ch2) at (2.5,-1.75) {\small{channel}};

\draw[latex-latex] (HTR1.south) -- (LTR1.north);
\draw[latex-latex] (HTR2.south) -- (LTR2.north);
\draw[latex-latex] (ch.east) -- (LTR2.west);
\draw[latex-latex] (ch.west) -- (LTR1.east);

\draw[-,dashed,line width=1.2pt] (-5,-0.25) -- (7,-0.25);

\node (hi) at (-3,0.1) {\small{High Layer (HL)}};
\node (lo) at (-3,-0.7) {\small{Low Layer (LL)}};

\end{tikzpicture}

%% file: figures/cascade_a.tex
\begin{tikzpicture}

\node[rectangle,draw,minimum width=2.7cm,minimum height=1cm] (node1) at (0,0) {\small Computation};
\node[rectangle,draw,minimum width=2.7cm,minimum height=1cm] (node2) at (4,0) {\small Compression};
\node[rectangle,draw,minimum width=2.7cm,minimum height=1cm] (node3) at (8,0) {\small Communication};
\draw[->] (-3,0) -- node[above,midway] {\small $D_1$} (node1.west);
\draw[->] (node1.east) -- node[above,midway] {\small $D_2$} (node2.west);
\draw[->] (node2.east) -- node[above,midway] {\small $D_3$} (node3.west);
\draw[->] (node3.east) -- node[above,midway] {\small $D_4$} (11,0);
\node at (0,1.5){};

\end{tikzpicture}

%% file: figures/cascade_b.tex
\begin{tikzpicture}

\node[rectangle,draw,minimum width=2.7cm,minimum height=1cm] (node1) at (0,0) {\small Computation};
\node[rectangle,draw,minimum width=2.7cm,minimum height=1cm] (node2) at (4,0) {\small Compression};
\node[rectangle,draw,minimum width=2.7cm,minimum height=1cm] (node3) at (8,0) {\small Communication};
\draw[->] (-3,0) -- node[above,midway] {\small $D_1$} (node1.west);
\draw[->] (node1.east) -- node[above,midway] {\small $D_2$} (node2.west);
\draw[->] (node2.east) -- node[above,midway] {\small $D_3$} (node3.west);
\draw[->] (node3.east) -- node[above,midway] {\small $D_4$} (11,0);
\draw[<->,red] (node1.north) -- (0,1) -- node[above,midway] {\small Metadata} (4,1) -- (node2.north);
\node at (0,2) {};

\end{tikzpicture}

%% file: figures/cascade_c.tex
\begin{tikzpicture}

\node[rectangle,draw,minimum width=6.7cm,minimum height=1cm] (node1) at (2,0) {\small Computation + compression};
\node[rectangle,draw,minimum width=2.7cm,minimum height=1cm] (node3) at (8,0) {\small Communication};
\draw[->] (-3,0) -- node[above,midway] {\small $D_1$} (node1.west);
\draw[->] (node1.east) -- node[above,midway] {\small $D_3$} (node3.west);
\draw[->] (node3.east) -- node[above,midway] {\small $D_4$} (11,0);
\node at (0,1.5){};

\end{tikzpicture}

%% file: figures/consensus.tex
\begin{tikzpicture}[>=latex]

\node[inner sep=0pt] (u) at (0,2.5) {\scalebox{-1}[1]{\includegraphics[width=.05\textwidth]{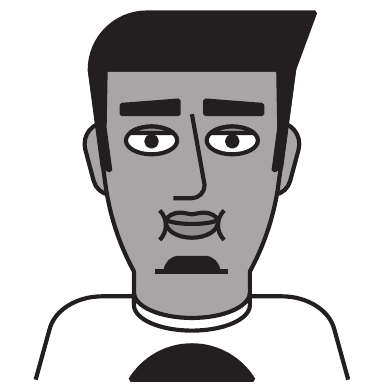}}};
\node[inner sep=0pt] (v) at (1,2.5) {\scalebox{-1}[1]{\includegraphics[width=.05\textwidth]{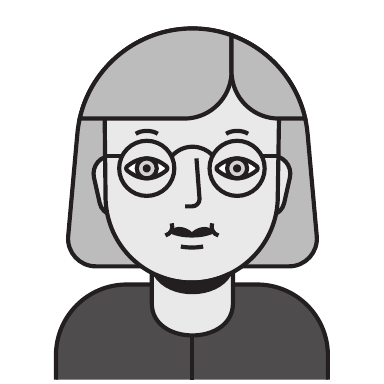}}};
\node[inner sep=0pt] (w) at (2,2.5) {\scalebox{-1}[1]{\includegraphics[width=.05\textwidth]{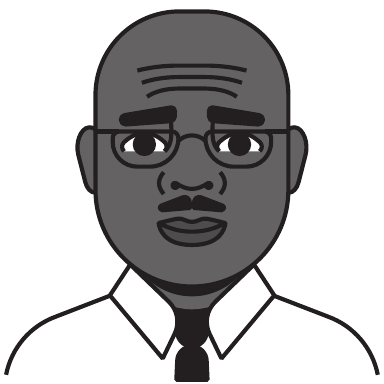}}};
\node[inner sep=0pt] (x) at (3,2.5) {\scalebox{-1}[1]{\includegraphics[width=.05\textwidth]{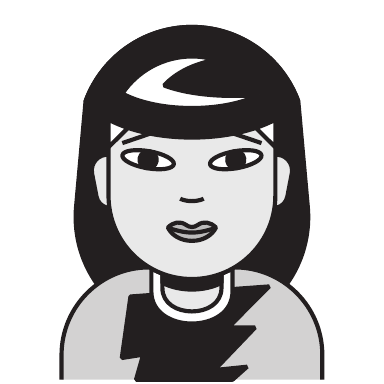}}};
\node[inner sep=0pt] (y) at (4,2.5) {\scalebox{-1}[1]{\includegraphics[width=.05\textwidth]{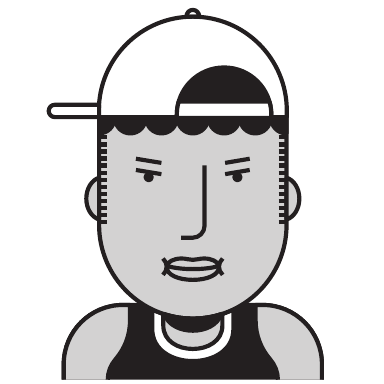}}};
\node[inner sep=0pt] (z) at (5,2.5) {\scalebox{-1}[1]{\includegraphics[width=.05\textwidth]{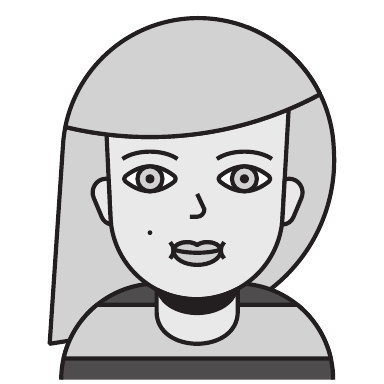}}};

\node[circle,fill=red,radius=0.3cm] at (0, 1.7){};
\node[circle,fill=blue,radius=0.3cm] at (1, 1.7){};
\node[circle,fill=blue,radius=0.3cm] at (2, 1.7){};
\node[circle,fill=red,radius=0.3cm] at (3, 1.7){};
\node[circle,fill=red,radius=0.3cm] at (4, 1.7){};
\node[circle,fill=blue,radius=0.3cm] at (5, 1.7){};

\node at (2.5,4) {Communication round 1};
\node at (10.5,4) {Communication round 2};
\node at (2.5,0.5) {Communication round 3};
\node at (10.5,0.5) {Communication round 4};

\draw[<->,green] ([xshift=-0.1cm]x.east) -- ([xshift=0.1cm]y.west);
\draw[<->,green] ([xshift=-0.1cm]y.east) -- ([xshift=0.1cm]z.west);
\draw[<->,green] ([xshift=-0.1cm]v.east) -- ([xshift=0.1cm]w.west);
\draw[<->,green] (u.north) to [out=30,in=150](z.north);

\node[inner sep=0pt] (u2) at (8,2.5) {\scalebox{-1}[1]{\includegraphics[width=.05\textwidth]{figures/Umer_front.pdf}}};
\node[inner sep=0pt] (v2) at (9,2.5) {\scalebox{-1}[1]{\includegraphics[width=.05\textwidth]{figures/Victoria_front.pdf}}};
\node[inner sep=0pt] (w2) at (10,2.5) {\scalebox{-1}[1]{\includegraphics[width=.05\textwidth]{figures/Walt_front.pdf}}};
\node[inner sep=0pt] (x2) at (11,2.5) {\scalebox{-1}[1]{\includegraphics[width=.05\textwidth]{figures/Xia_front.pdf}}};
\node[inner sep=0pt] (y2) at (12,2.5) {\scalebox{-1}[1]{\includegraphics[width=.05\textwidth]{figures/Yoshi_front.pdf}}};
\node[inner sep=0pt] (z2) at (13,2.5) {\scalebox{-1}[1]{\includegraphics[width=.05\textwidth]{figures/Zoya_front.pdf}}};

\node[circle,fill=red,radius=0.3cm] at (8, 1.7){};
\node[circle,fill=blue,radius=0.3cm] at (9, 1.7){};
\node[circle,fill=blue,radius=0.3cm] at (10, 1.7){};
\node[circle,fill=red,radius=0.3cm] at (11, 1.7){};
\node[circle,fill=red,radius=0.3cm] at (12, 1.7){};
\node[circle,fill=red,radius=0.3cm] at (13, 1.7){};

\draw[<->,green] ([xshift=-0.1cm]u2.east) -- ([xshift=0.1cm]v2.west);
\draw[<->,green] ([xshift=-0.1cm]w2.east) -- ([xshift=0.1cm]x2.west);
\draw[<->,green] (w2.north) to [out=30,in=150](z2.north);
\draw[<->,green] (u2.north) to [out=30,in=150](x2.north);

\node[inner sep=0pt] (u3) at (0,-1) {\scalebox{-1}[1]{\includegraphics[width=.05\textwidth]{figures/Umer_front.pdf}}};
\node[inner sep=0pt] (v3) at (1,-1) {\scalebox{-1}[1]{\includegraphics[width=.05\textwidth]{figures/Victoria_front.pdf}}};
\node[inner sep=0pt] (w3) at (2,-1) {\scalebox{-1}[1]{\includegraphics[width=.05\textwidth]{figures/Walt_front.pdf}}};
\node[inner sep=0pt] (x3) at (3,-1) {\scalebox{-1}[1]{\includegraphics[width=.05\textwidth]{figures/Xia_front.pdf}}};
\node[inner sep=0pt] (y3) at (4,-1) {\scalebox{-1}[1]{\includegraphics[width=.05\textwidth]{figures/Yoshi_front.pdf}}};
\node[inner sep=0pt] (z3) at (5,-1) {\scalebox{-1}[1]{\includegraphics[width=.05\textwidth]{figures/Zoya_front.pdf}}};

\node[circle,fill=red,radius=0.3cm] at (0, -1.8){};
\node[circle,fill=blue,radius=0.3cm] at (1, -1.8){};
\node[circle,fill=red,radius=0.3cm] at (2, -1.8){};
\node[circle,fill=red,radius=0.3cm] at (3, -1.8){};
\node[circle,fill=red,radius=0.3cm] at (4, -1.8){};
\node[circle,fill=red,radius=0.3cm] at (5, -1.8){};

\draw[<->,green] (u3.north) to [out=30,in=150](w3.north);
\draw[<->,green] (v3.north) to [out=30,in=150](z3.north);
\draw[<->,green] (v3.north) to [out=30,in=150](x3.north);

\node[inner sep=0pt] (u4) at (8,-1) {\scalebox{-1}[1]{\includegraphics[width=.05\textwidth]{figures/Umer_front.pdf}}};
\node[inner sep=0pt] (v4) at (9,-1) {\scalebox{-1}[1]{\includegraphics[width=.05\textwidth]{figures/Victoria_front.pdf}}};
\node[inner sep=0pt] (w4) at (10,-1) {\scalebox{-1}[1]{\includegraphics[width=.05\textwidth]{figures/Walt_front.pdf}}};
\node[inner sep=0pt] (x4) at (11,-1) {\scalebox{-1}[1]{\includegraphics[width=.05\textwidth]{figures/Xia_front.pdf}}};
\node[inner sep=0pt] (y4) at (12,-1) {\scalebox{-1}[1]{\includegraphics[width=.05\textwidth]{figures/Yoshi_front.pdf}}};
\node[inner sep=0pt] (z4) at (13,-1) {\scalebox{-1}[1]{\includegraphics[width=.05\textwidth]{figures/Zoya_front.pdf}}};

\node[circle,fill=red,radius=0.3cm] at (8, -1.8){};
\node[circle,fill=red,radius=0.3cm] at (9, -1.8){};
\node[circle,fill=red,radius=0.3cm] at (10, -1.8){};
\node[circle,fill=red,radius=0.3cm] at (11, -1.8){};
\node[circle,fill=red,radius=0.3cm] at (12, -1.8){};
\node[circle,fill=red,radius=0.3cm] at (13, -1.8){};

\end{tikzpicture}

%% file: figures/model_update.tex
\begin{tikzpicture}[>=latex]

\node[rectangle,draw,minimum width=3.5cm,minimum height=0.7cm] (round1) at (-1.5,4.5) {\scriptsize{Global model broadcasting}};
\node[rectangle,draw,minimum width=3.5cm,minimum height=0.7cm] (round1) at (2,4.5) {\scriptsize{Local gradient computation}};
\node[rectangle,draw,minimum width=3.5cm,minimum height=0.7cm] (round1) at (5.5,4.5) {};

\node (lca1) at (5.5,4.6) {\scriptsize{Local gradient computation}};
\node (lca2) at (5.5,4.3) {\scriptsize{for global model updating}};

\draw[-latex] (-3.5,4.15) --  (8,4.15);
\draw[-] (-3.25,4.15) --  (-3.25,3.8);
\draw[-] (7.25,4.15) --  (7.25,3.8);
\draw[latex-latex] (-3.25,3.8) -- node [midway,fill=white]{\scriptsize{Single communication round}} (7.25,3.8);

\node[inner sep=0pt] (bs) at (0,0) {\includegraphics[width=.1\textwidth]{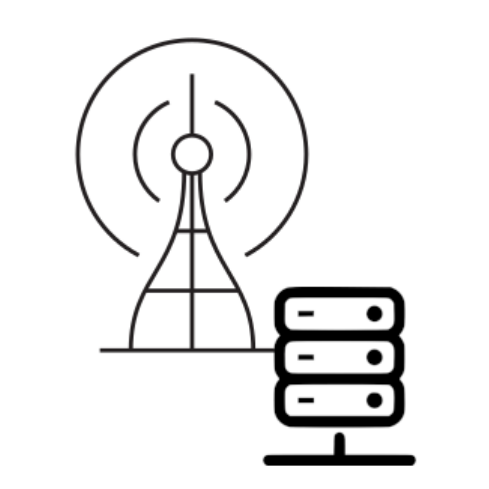}};
\node[inner sep=0pt] (usr1) at (3,1.5) {\scalebox{-1}[1]{\includegraphics[width=.05\textwidth]{figures/Umer_mobile.pdf}}};
\node[inner sep=0pt] (usr2) at (3,0) {\scalebox{-1}[1]{\includegraphics[width=.05\textwidth]{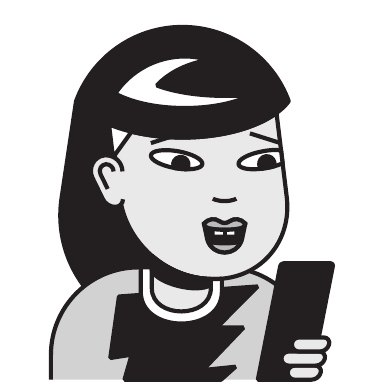}}};
\node[inner sep=0pt] (usr3) at (3,-1.5) {\scalebox{-1}[1]{\includegraphics[width=.05\textwidth]{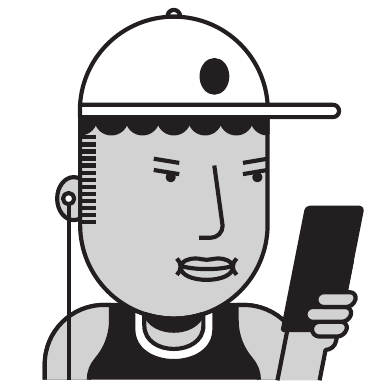}}};

\node[inner sep=0pt] (mem1) at (7,1.5) {\includegraphics[width=.075\textwidth]{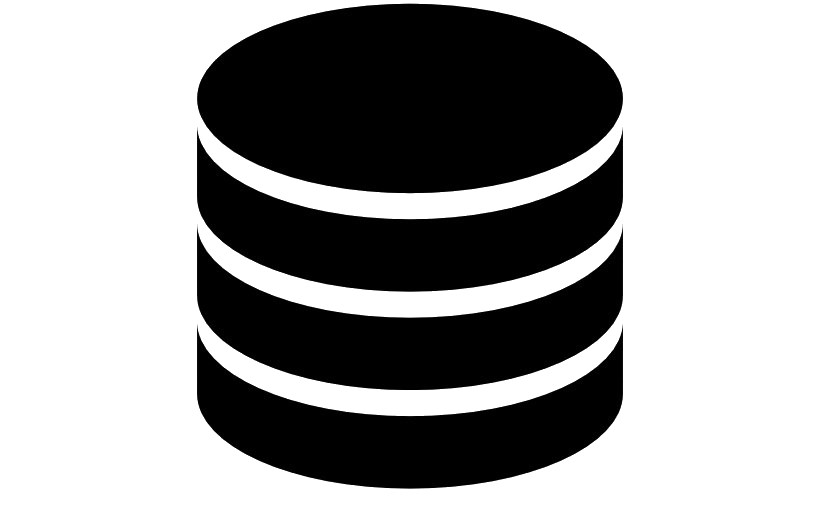}};
\node[inner sep=0pt] (mem3) at (7,-1.5) {\includegraphics[width=.075\textwidth]{figures/database}};

\node[rectangle,draw,minimum width=1.5cm,minimum height=1cm] (comp1) at (5,1.5) {};
\node[rectangle,draw,minimum width=1.5cm,minimum height=1cm] (mod1) at (5,3) {};
\node[rectangle,draw,minimum width=1.5cm,minimum height=1cm] (comp2) at (5,-1.5) {};
\node[rectangle,draw,minimum width=1.5cm,minimum height=1cm] (mod2) at (5,-3) {};

\draw[latex-] ([yshift=8pt]bs.east) --  ([yshift=1.5pt]usr1.west);
\draw[-latex,dotted] ([yshift=5pt]bs.east) --  ([yshift=-1.5pt]usr1.west);
\draw[latex-] ([yshift=1.5pt]bs.east) --  ([yshift=1.5pt]usr2.west);
\draw[-latex,dotted] ([yshift=-1.5pt]bs.east) --  ([yshift=-1.5pt]usr2.west);
\draw[latex-] ([yshift=-5pt]bs.east) --  ([yshift=1.5pt]usr3.west);
\draw[-latex,dotted] ([yshift=-8pt]bs.east) --  ([yshift=-1.5pt]usr3.west);

\node (n1) at (6.5,1.5) {};
\node (n2) at (6.5,3) {};
\node (n3) at (3,3) {};

\draw[latex-] (usr1.east) --  (comp1.west);
\draw[latex-] (usr3.east) --  (comp2.west);
\draw[latex-] (comp1.east) --  ([xshift=5pt]mem1.west);
\draw[latex-] (comp2.east) --  ([xshift=5pt]mem3.west);
\draw[-] (mod1.east) --  (6.2,3);
\draw[-] ([yshift=3pt]usr1.north) --  (3,3);
\draw[latex-] (mod1.west) --  (3,3);
\draw[latex-] (6.2,1.5) --  (6.2,3);
\draw[-] (mod2.east) --  (6.2,-3);
\draw[-] ([yshift=-3pt]usr3.south) --  (3,-3);
\draw[latex-] (mod2.west) --  (3,-3);
\draw[latex-] (6.2,-1.5) --  (6.2,-3);

\draw(0.3,-1.2) -- (0.3,-2.4) -- (-1.2,-2.4) -- (-1.2,-1.4) -- (0.1,-1.4) -- cycle;
\draw(-2.5,0) -- (-2.5,1) -- (-1,1) -- (-1,0.2) -- (-0.8,0) -- cycle;

\node (ga1) at (-0.45,-1.75) {\scriptsize{Gradient}};
\node (ga2) at (-0.45,-2.05) {\scriptsize{aggregation}};

\node (gm1) at (-1.75,0.8) {\scriptsize{Global}};
\node (gm2) at (-1.75,0.5) {\scriptsize{model}};
\node (gm3) at (-1.75,0.2) {\scriptsize{update}};

\draw[-] (-1.2,-1.9) --  (-1.75,-1.9);
\draw[latex-] (-1.75,0) --  (-1.75,-1.9);

\node (es) at (0,-1) {\scriptsize{Edge server}};
\node (up1) at (1.5,1.75) {\scriptsize{Local gradient}};
\node (up2) at (1.5,1.45) {\scriptsize{uploading}};
\node (bc1) at (1.5,-1.45) {\scriptsize{Global model}};
\node (bc2) at (1.5,-1.75) {\scriptsize{broadcasting}};
\node (g1) at (5,1.65) {\scriptsize{Gradient}};
\node (g2) at (5,1.35) {\scriptsize{computation}};
\node (m1) at (5,3.15) {\scriptsize{Global}};
\node (m2) at (5,2.85) {\scriptsize{model}};
\node (g3) at (5,-1.35) {\scriptsize{Gradient}};
\node (g4) at (5,-1.65) {\scriptsize{computation}};
\node (m1) at (5,-2.85) {\scriptsize{Global}};
\node (m2) at (5,-3.15) {\scriptsize{model}};
\node (d1) at (7,2.35) {\scriptsize{Local}};
\node (d2) at (7,2.05) {\scriptsize{dataset}};
\node (d3) at (7,-0.65) {\scriptsize{Local}};
\node (d4) at (7,-0.95) {\scriptsize{dataset}};

\end{tikzpicture}

%% file: figures/edge_update.tex
\begin{tikzpicture}[>=latex]

\node[rectangle,draw,fill=gray!30!white,minimum width=4cm,minimum height=2cm] (ed) at (0,1) {};
\node[rectangle,draw,fill=gray!30!white,minimum width=4cm,minimum height=2cm] (es) at (9,1) {};
\node[inner sep=0pt] (bs1) at (3,1.5) {\includegraphics[width=.1\textwidth]{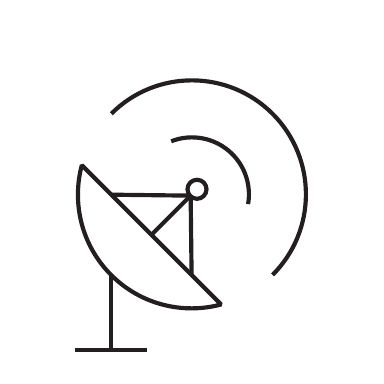}};
\node[inner sep=0pt] (bs2) at (6,1.5) {\scalebox{-1}[1]{\includegraphics[width=.1\textwidth]{figures/antenna.pdf}}};

\node[rectangle,draw,fill=white,minimum width=1.5cm,minimum height=1cm] (fe) at (-1,1) {};
\node[rectangle,draw,fill=white,minimum width=1.5cm,minimum height=1cm] (c) at (1,1) {\scriptsize{Coder}};
 \draw[-latex] (fe.east) --  (c.west);
 \draw[-latex] (-2.3,1) --  (fe.west);
 \draw[latex-] (2.5,1) --  (c.east);
\node(il) at (0,0.25){\scriptsize{Inferred labels}}; 
\node(dev) at (0,1.7){\small\textbf{Edge Device}}; 
 
 \draw[-latex] (2.5,0.9) -- (2.3,0.9) -- (2.3,0.25) -- (il.east);

\node[rectangle,draw,fill=white,minimum width=1.5cm,minimum height=1cm] (d) at (8,1) {\scriptsize{Decoder}};
\node[rectangle,draw,fill=white,minimum width=1.5cm,minimum height=1cm] (cl) at (10,1) {\scriptsize{Classifier}};
\draw[-latex] (d.east) --  (cl.west);
\draw[-latex] (6.5,1) --  (d.west);
\node(dev) at (9,1.7){\small\textbf{Edge Server}}; 
\draw[-latex] (cl.south) --  (10,0.25) -- node[midway,fill=gray!30!white]{\scriptsize{Inferred labels}} (6.7,0.25) -- (6.7,0.9) -- (6.5,0.9);

 \draw[-latex,dashed] ([yshift=3mm]bs1.east) -- node[midway,above]{\scriptsize{Feature vectors}} ([yshift=3mm]bs2.west);
 \draw[latex-,dotted] ([yshift=-1mm]bs1.east) -- node[midway,below]{\scriptsize{Inferred features}} ([yshift=-1mm]bs2.west);
 
\node (fe1) at (-1,1.15) {\scriptsize{Feature}};
\node (fe2) at (-1,0.85) {\scriptsize{extraction}};
\node (d) at (-2.8,1.15) {\scriptsize{Data}};
\node (d) at (-2.8,0.85) {\scriptsize{samples}};

\end{tikzpicture}